\documentclass{ctr}

\usepackage{ctrfont}
\usepackage{natbib}

\usepackage{url}
\usepackage{amsmath}
\usepackage{amssymb}
\usepackage{soul}
\usepackage{subfig}
\usepackage{color}
\usepackage[normalem]{ulem}

\usepackage[labelsep=period]{caption}
\renewcommand{\tablename}{\sc Table}
\renewcommand{\figurename}{\sc Figure}

\usepackage{graphicx}
\usepackage{psfrag}
\usepackage{epsfig}

\newcommand{\p}{\partial}

\newcommand{\damp}{\mu}
\newcounter{saveequation}
\newcommand {\bu} {{\mbox {\boldmath $u $}}}
\newcommand {\bq} {{\mbox {\boldmath $q $}}}
\newcommand{\e}{\varepsilon}
\newcommand{\bse}{\begin{subequations}}
\newcommand{\ese}{\end{subequations}}
\newcommand{\beq}{\begin{equation}}
\newcommand{\eeq}{\end{equation}}

\newcommand {\ba} {\begin {array}}
\newcommand {\ea} {\end {array}}

\title{Wall turbulence with constrained energy extraction from the mean flow}
\shorttitle{Wall turbulence with constrained energy extraction}
\author{A. Lozano-Dur\'an, M. Karp, 
\and N.~C. Constantinou\footnote{Research School of Earth Sciences, Australian National University, Australia}\footnote{ARC Centre of Excellence for Climate Extremes, Australian National University, Australia}
}
\shortauthor{Lozano-Dur\'an et~al.}

\begin{document}


\maketitle


\section{Motivation and objectives} 

Turbulence is the primary example of a highly nonlinear
phenomenon. However, there is evidence that some processes of shear
turbulence are controlled by linear dynamics, in particular the
mechanism by which energy is transferred from the mean velocity
component of the flow to the spatially and temporally evolving
perturbations \citep[e.g.,][]{Farrell-Ioannou-1998a, Kim2000,
  Jimenez2013a}.  The goal of the present work is to investigate the
mechanism dominating the energy transfer from the mean flow to the
fluctuating field in wall-bounded turbulence.

It is agreed that the streamwise rolls and streaks are ubiquitous in
wall-shear flow \citep{Klebanoff1962, Kline1967} and that they are
involved in a quasi-periodic regeneration cycle \citep{Panton2001,
  Adrian2007, Smits2011, Jimenez2012, Jimenez2018}.  The space-time
structure of rolls and streaks is believed to play an important role
in sustaining and carrying shear-driven turbulence
\citep[e.g.,][]{Kim1971, Jimenez1991, Hamilton1995, Waleffe1997,
  Schoppa2002, Jimenez2012}. The ultimate cause maintaining this
self-sustaining cycle, and hence turbulence, is the energy extraction
from the flow mean shear. Within the fluid mechanics community, there
have been several mechanisms proposed as plausible scenarios for how
this energy extraction occurs. Conceptually, we can divide these
mechanisms into three categories: (i)~modal inflectional instability
of the mean cross-flow, (ii)~non-modal transient growth, and
(iii)~non-modal transient growth assisted by parametric instability of
the time-varying mean cross-flow.

In the first mechanism, it is hypothesized that the energy is
transferred from the cross-flow mean profile $U(y,z,t)$ ($y$ and $z$
are the wall-normal and spanwise directions, respectively) to the flow
fluctuations through a modal inflectional instability
\citep{Waleffe1997} in the form of a corrugated vortex sheet
\citep{Kawahara2003} or of intense localized patches of low-momentum
fluid \citep{Hack2018}. The second mechanism involves the collection
of fluid near the wall by streamwise vortices that is subsequently
organized into streaks via the lift-up mechanism \citep{Landahl1975,
  Butler1992, Jimenez2012}.  In this case, the mean flow, while
modally stable, it is able to support the growth of perturbations for
a transient time owing to the non-normality of the linear operator that
governs the evolution of fluctuations. This process is referred to as
non-modal transient growth \citep[e.g.,][]{Schmid2007}. Additional
studies suggest that the generation of streaks are due to the
structure-forming properties of the linearized Navier--Stokes
operator, independent of any organized vortices
\citep{Chernyshenko2005}, but the non-modal transient growth is still
invoked. The transient growth scenario gained even more popularity
since the work by \citet{Schoppa2002}, who argued that transient growth
may be the most relevant mechanism not only for streak formation but
also for their eventual breakdown.  \citet{Schoppa2002} showed that
most streaks detected in actual wall-turbulence simulations are indeed
modally stable. Instead, the loss of stability of the streaks is
better explained by transient growth of perturbations that leads to
vorticity sheet formation and nonlinear saturation. Finally, a third
mechanism has been proposed in recent years by \citet{Farrell2012}
and \citet{Farrell2016}. Farrell and co-workers adopted the
perspective of statistical state dynamics (SSD) to develop a theory
for the maintenance of wall turbulence. Through the SSD framework, it
is revealed that the perturbations are maintained by an essentially
time-dependent, parametric, non-normal interaction with the streak,
rather than by the inflectional instability of the streaky flow
discussed above \citep[see also][]{Farrell2017}.

The three different mechanisms, each capable of leading to the
observed turbulence structure, are rooted in theoretical or conceptual
arguments. Whether the energy transfer from the mean cross-flow to
fluctuations in wall-bounded turbulence occurs through any or a
combination of these mechanisms remains unclear. Most of the theories
stem from linear stability theory, which has proven very successful in
providing a theoretical framework to explain the lengths and time
scales observed in the flow. However, an appropriate base flow for the
linearization must be selected \emph{a-priori} depending on the flow
state of interest; this introduces some degree of
arbitrariness. Moreover, quantitative results are known to be
sensitive to the details of the base state \citep{Vaughan2011}.  For
example, there have been considerable efforts to explain and control
turbulent structure and length scales by linearizing around the
turbulent mean profile obtained by averaging in homogeneous directions
and time
\citep[e.g.,][]{Hogberg2003,DelAlamo2006,Hwang2010b}. However, the
turbulent mean profile is known to be always modally stable, and thus
mechanisms~(i) and~(iii) are precluded. The self-sustained turbulent
state is intimately related to the roll--streak structure
\citep[e.g.,][]{Waleffe1997}, and this suggests that the rolls--streaks
should be part of the base flow, as pointed out by the SSD theory.

Another criticism of linear studies is that turbulence is a highly
nonlinear phenomenon, and a full self-sustained cycle cannot be
uncovered from a single set of linearized equations. For example, in
turbulent channel flows, the classic linearization around the mean
velocity profile does not account for the redistribution of energy
from the streamwise velocity component to the cross-flow, which is the
prevailing energy transfer on average \citep{Mansour1988}. In order to
capture different energy transfer mechanisms, the base state for
linearization should be selected accordingly.  In this regard,
eigenmodes or optimal solutions should not be taken as representative
of the actual flow and, if they are considered valid, the time and
length scales for which linearization remains meaningful become
relevant issues that are barely discussed in the literature.

Here, we attempt to assess the relative importance of the three
proposed mechanisms for energy extraction from the mean flow in wall
turbulence. For now, we mainly focus on whether we can obtain a
self-sustained turbulent-like flow when a particular mechanism is
inhibited. First, we present some diagnostics from direct numerical
simulations of wall turbulence. Second, we designed three numerical
experiments each of which is dominated by the energy extraction from
modal instability, non-modal transient growth, or transient growth
with parametric instability.  The proposed experiments are fully
nonlinear systems to close the feedback loop between mean cross-flow
and perturbations, enabling in this manner the possibility of
sustained turbulence. The experiments are accompanied by some
preliminary results.

The Brief is organized as follows: Section~\ref{sec:numerical}
contains the numerical details of the simulations and the stability
analysis of the mean cross-flow.  The results are presented in
Section~\ref{sec:results}, which is further subdivided into three
subsections describing the details of the flow set-up and the
corresponding results. Finally, conclusions and future directions are
offered in Section~\ref{sec:conclusions}.

\section{Numerical experiments of turbulent channel flow}\label{sec:numerical}

\subsection{Numerical setup}\label{subsec:numerical}

The baseline case is a plane turbulent channel flow at $Re_\tau=184$,
with streamwise, wall-normal, and spanwise domain sizes equal to $L_x^+
\approx 337$, $L_y^+ \approx 368$, and $L_z^+ \approx 168$,
respectively, where~$+$ denotes wall units defined in terms of the
kinematic viscosity $\nu$ and friction velocity at the wall
$u_\tau$. The channel half-height is denoted by $h$.
\citet{Jimenez1991} showed that simulations in this domain constitute
an elemental structural unit containing a single streamwise streak and
a pair of staggered quasi-streamwise vortices, which reproduce fairly
well the statistics of the flow in larger domains.  We refer to this
case as CH180.

We consider three additional numerical set-ups by solving
\begin{eqnarray}
\label{eq:NS}
\frac{\partial u_i}{\partial t} = -\frac{\partial u_i u_j}{\partial
  x_j} - \frac{\partial p}{\partial x_i} + \nu \frac{\partial^2 u_i
}{\partial x_k \partial x_k} + f_i, \quad \frac{\partial u_i}{\partial
  x_i} = 0,
\end{eqnarray}
where repeated indices imply summation, $(u_1,u_2,u_3)=(u,v,w)$ are
streamwise, wall-normal, and spanwise velocities with respective
coordinates $(x_1,x_2,x_3)=(x,y,z)$, $p$ is the pressure, and
$f_i=f_i(x,y,z,t)$ is a forcing term aiming to prevent one or several
of the proposed energy injection mechanisms. The functional form of
$f_i$ is discussed below for each particular case.

The simulations are performed with a staggered, second-order, finite
differences scheme \citep{Orlandi2000} and a fractional-step method
\citep{Kim1985} with a third-order Runge-Kutta time-advancing scheme
\citep{Wray1990}.  The solution is advanced in time using a constant
time step such that the Courant--Friedrichs--Lewy condition is below
0.5.  The streamwise and spanwise resolutions are $\Delta x^+\approx
6.5$ and $\Delta z^+\approx3.3$, respectively, and the minimum and
maximum wall-normal resolutions are $\Delta
y_{\mathrm{min}}^+\approx0.2$ and $\Delta
y_{\mathrm{max}}^+\approx6.1$.  All the simulations were run for at
least $100h/u_\tau$ after transients. The code has been validated in
previous studies in turbulent channel flows \citep{Lozano2016_Brief,
  Bae2018b, Bae2018}, and flat-plate boundary layers
\citep{Lozano2018}.

We introduce the averaging operators $\langle \,\cdot\, \rangle_{x}$,
$\langle \,\cdot\, \rangle_{xz}$, and $\langle \,\cdot\,
\rangle_{xzt}$ which denote averaging in $x$ direction, $x$ and $z$
directions, and $x$, $z$ and $t$, respectively. The mean velocity
profile is defined as $\langle u \rangle_{xzt}$, the mean cross-flow
velocity profile as $U=\langle u \rangle_{x}$, and the fluctuating
velocities (or perturbations) as $u_1'= u_1 - U$, $u_2'=
u_2$, and $u_3'= u_3$.

\subsection{Linear stability of the mean cross-flow for case CH180}\label{subsec:CH180}

We investigate the stability of $A(U)$ that governs the linear
evolution of the fluctuating velocity $\boldsymbol{u}'=(u',v',w')$, i.e.,
\beq
        \frac{\partial \boldsymbol{u}'}{\partial t} = A(U) \boldsymbol{u}'.
\eeq
The analysis is performed for different times $t_0$ by assuming a
constant-in-time mean cross-flow $U(y,z,t_0)$. Occasionally, we
refer to the stability of operator $A(U)$ simply as the stability of $U$.
The details of the analysis are provided in the Appendix.

Figure~\ref{fig:CH180}(a) shows the time evolution of the maximum
growth rate of $A$ denoted by $\sigma_{\mathrm{max}}$ (largest real
part of the eigenvalues of $A$). The flow is modally unstable 90\% of
the time ($\sigma_{\mathrm{max}}>0$). Since we have assumed that $U$
does not evolve in time, it is pertinent to discuss the validity of
such an assumption. The time auto-correlation of $U$ is plotted in
Figure \ref{fig:CH180}(b), which reveals that 50\% and 100\%
de-correlation times are attained about $h/u_\tau$ and $4h/u_\tau$,
respectively.  Rigorously, only growth rates with characteristic times
$1/\sigma_{\mathrm{max}}$ much shorter than the characteristic
de-correlation time of the mean cross-flow should be taken as
representative of the linear stability of $U$. The results in Figure
\ref{fig:CH180} show that the flow is modally unstable 80\% of the
time history if we account for growth rates larger than $u_\tau/(4h)$,
and 40\% of the time for growth rates larger than $u_\tau/h$.  A
complementary metric to assess the validity of frozen-in-time $U$ is
the characteristic growth rate of $U$ defined as $\sigma_U =
(\mathrm{d}E_U/\mathrm{d}t)/(2E_U)$ with $E_U = \langle U^2/2
\rangle_{yz}$. The ratio $\sigma_{\mathrm{max}}/\sigma_U$ was found to
be on average $\approx 10$, i.e., the rate of change of $U$ is on
average ten times slower than the maximum growth rate predicted by
linear stability analysis. A tentative conclusion is that the
stability analysis of $U$ may not be quantitatively valid, but the
observed stability trends are probably correct and, hence, $U$
supports exponential growth of disturbances for a non-negligible
fraction of the flow history.
%
\begin{figure}
 \begin{center}
   \subfloat[][\phantom{a}]{\includegraphics[width=0.45\textwidth]{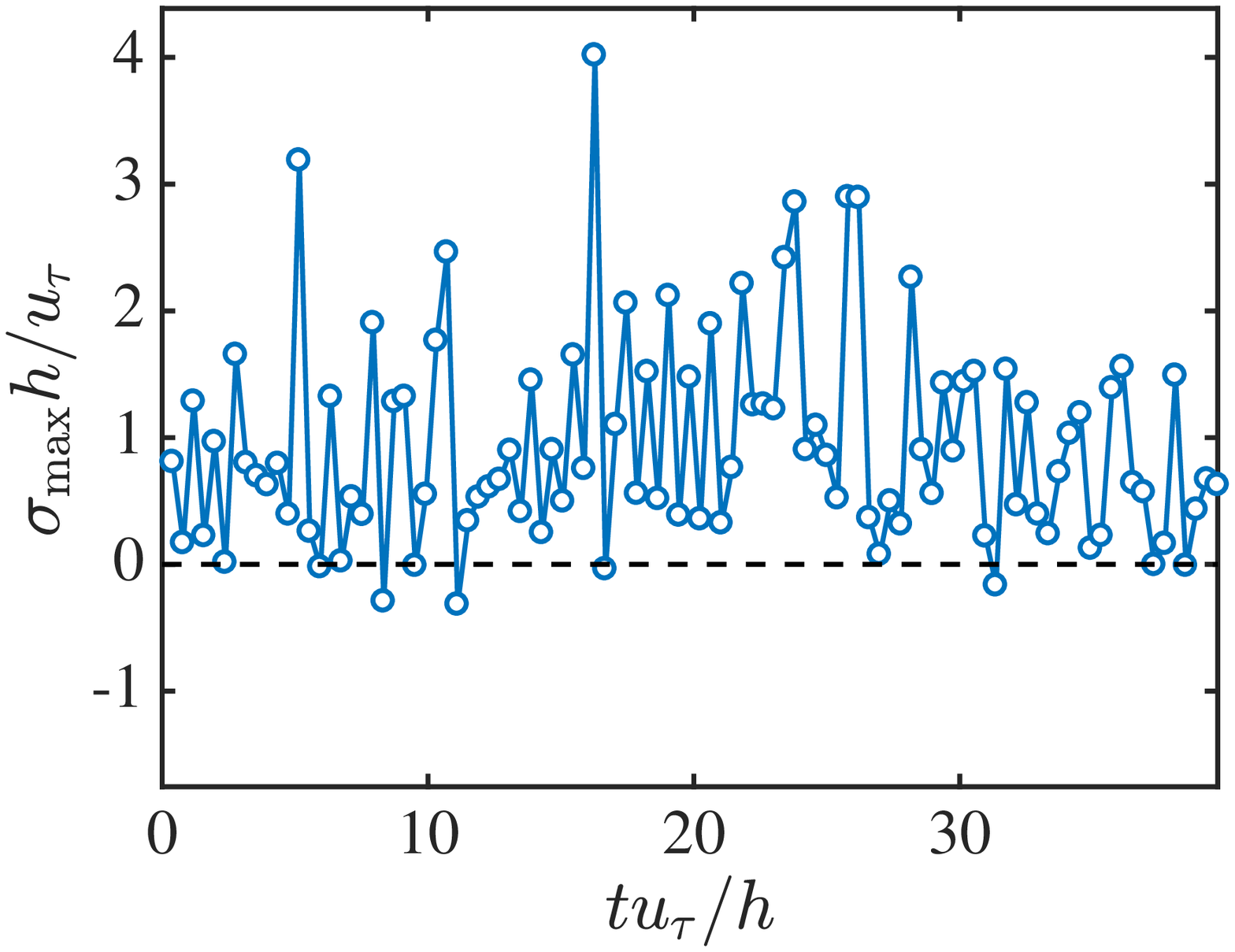} }
   \hspace{0.1cm}
   \subfloat[]{\includegraphics[width=0.50\textwidth]{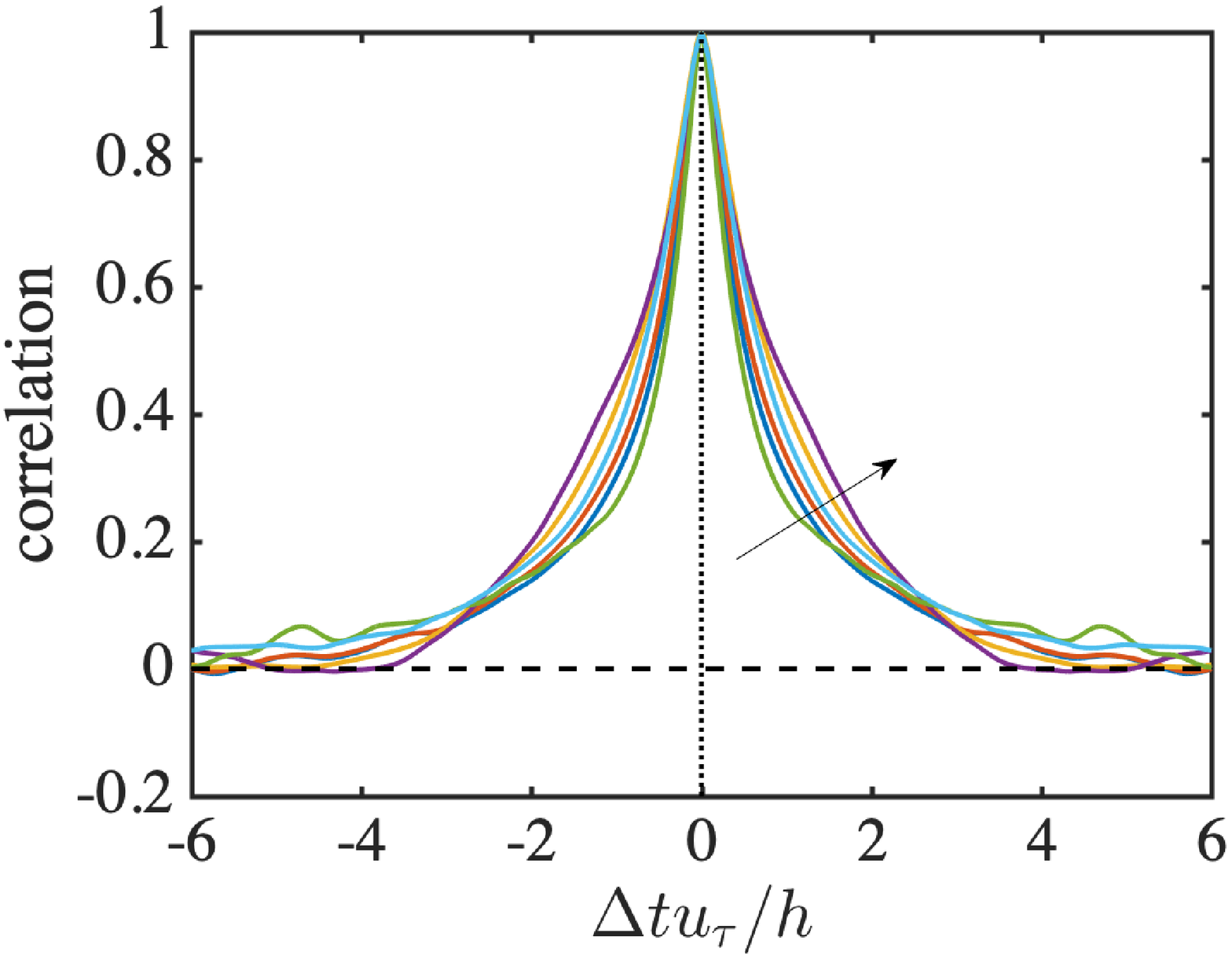} }
 \end{center}
\caption{(a) The evolution of the maximum growth rate of the mean
  cross-flow $U(y,z,t)$. (b) The auto-correlation of the cross-flow
  $\langle U(y,z,t) \rangle_z$. The different lines are for
  $y/h=0.01,0.04,0.10,0.22,0.45,0.80$. The arrow indicates increasing
  $y/h$. \label{fig:CH180}}
\end{figure}

\section{Experiments for discerning energy transfer mechanisms and preliminary results} \label{sec:results}

\subsection{Primary energy injection by modal instability}\label{subsec:numerical:modal}

\begin{figure}
 \vspace{0.2cm}
 \begin{center}
   \includegraphics[width=1\textwidth]{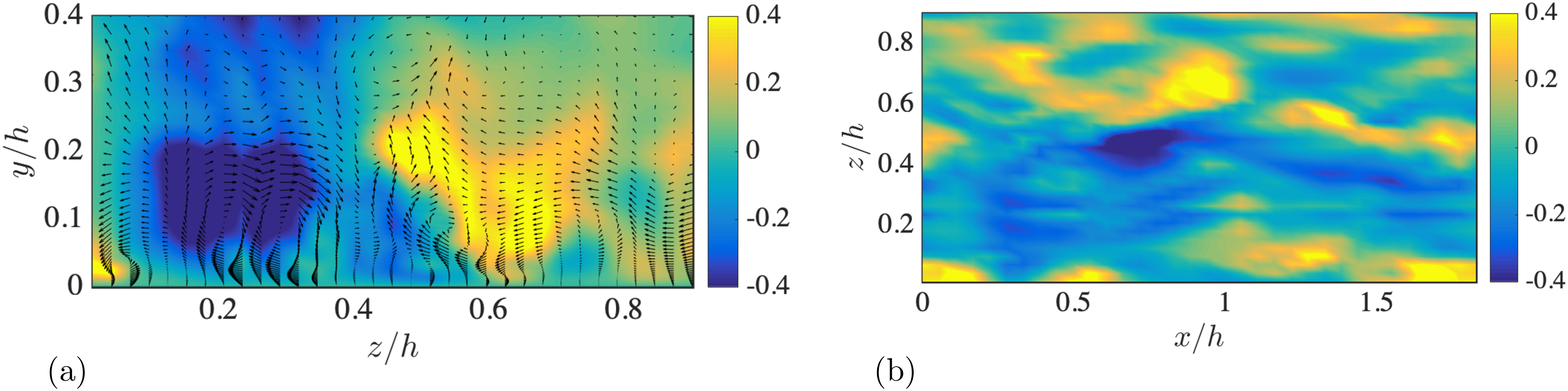}
 \end{center}
\caption{ Experiment with fixed, unstable $U$: (a) Instantaneous
  velocity field in a $z-y$ plane at $x=0h$. (b) Instantaneous
  streamwise velocity in a $x-z$ plane at $y=0.1h$. Colors represent
  streamwise velocity and arrows are cross-flow velocities. Velocities
  are scaled in wall units of the baseline case.
\label{fig:snap_modal}}
 \begin{center}
   \subfloat[]{ \includegraphics[width=0.45\textwidth]{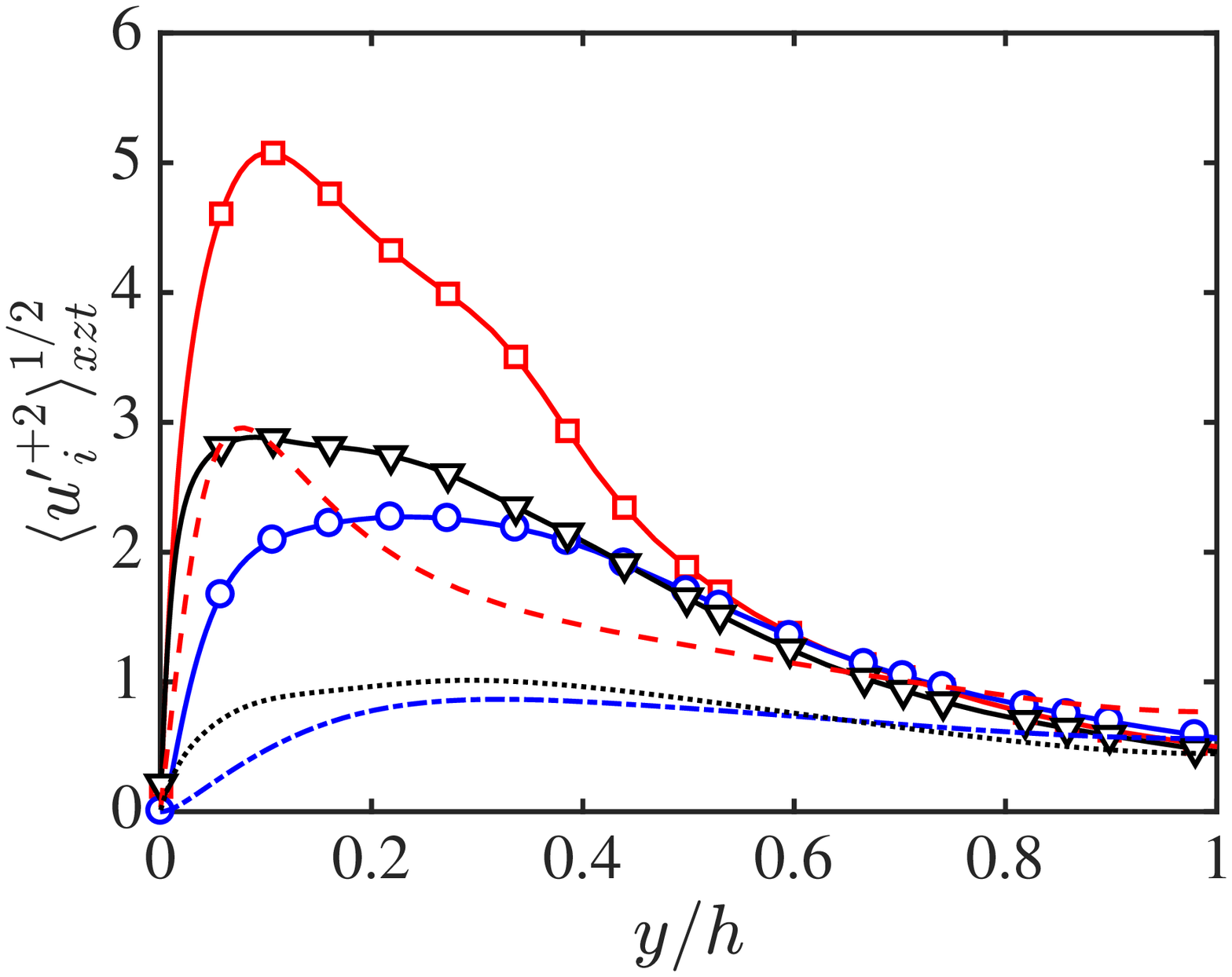} }
   \hspace{0.1cm}
   \subfloat[]{ \includegraphics[width=0.47\textwidth]{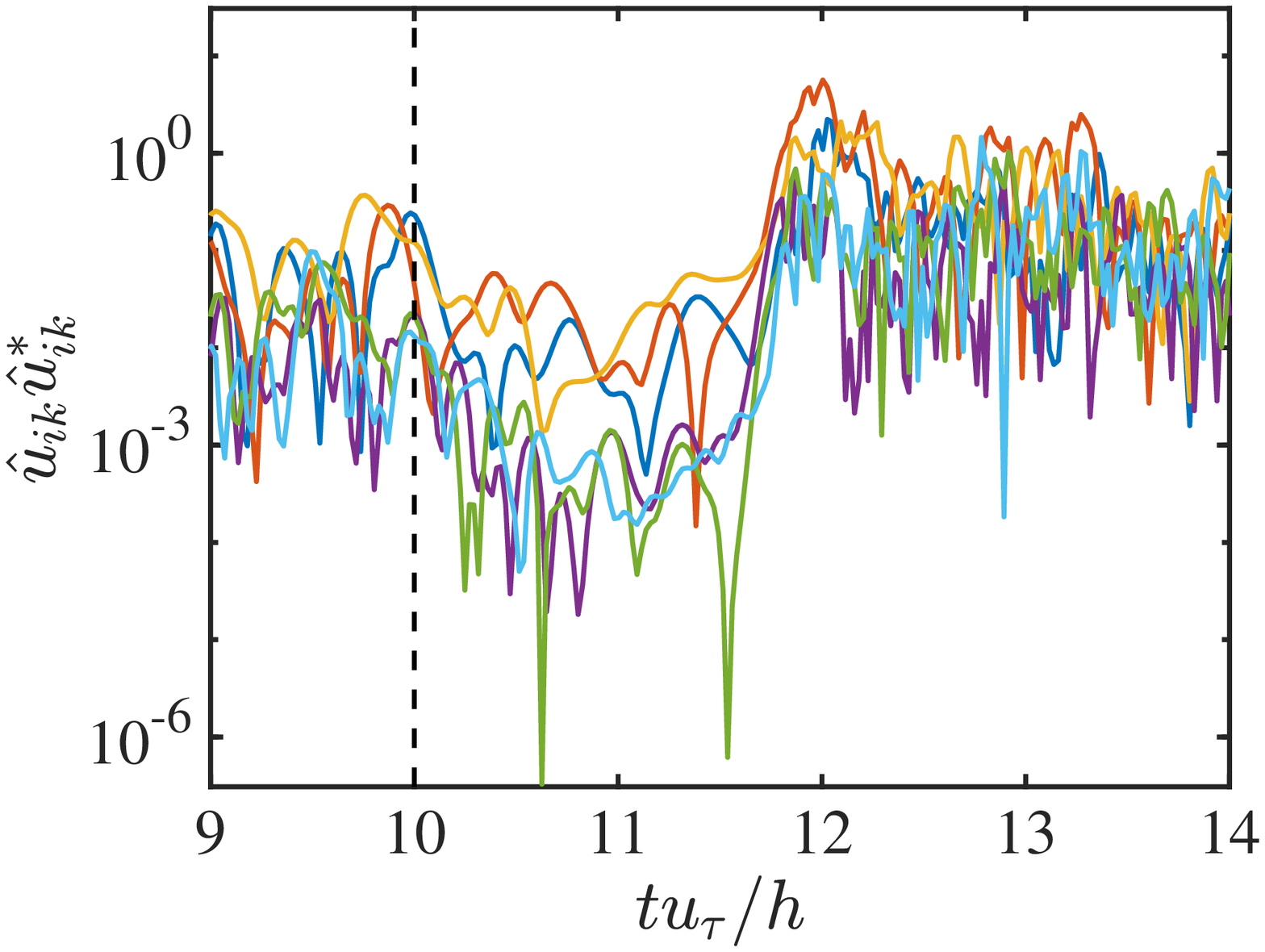} }
 \end{center}
\caption{ Experiment with fixed, unstable $U$: (a) Root-mean-squared
  fluctuating velocities for case CH180 (lines) and channel with
  frozen-in-time modally unstable mean cross-flow (symbols). Lines and
  symbols are: dashed and squares, streamwise; dash-dotted and
  circles, wall-normal; dotted and triangles, spanwise velocity
  fluctuations. (b) Time evolution of the energy associated with
  streamwise Fourier modes, $\hat u_{ik} \hat u^*_{ik}$, for $i=1,2,3$
  and $k=0,1,2$ at $y=0.1h$, where $*$ denotes complex
  conjugation. The mean cross-flow is frozen at $t u_\tau/h = 10$
  (dashed black line).
\label{fig:stats_modal}}
\end{figure}

The effect of modal instability is assessed by freezing in time the
mean cross-flow for case CH180 at time~$t_0$ when $U(y,z,t)$ is
modally unstable. At each time step, $f_1$ is computed such that
$U(y,z,t)=U(y,z,t_0)$ with $f_2=f_3=0$. Additionally, $\langle u
\rangle_{xzt}$ is set to the same value as in case CH180. The lack of
time evolution in $U$ eliminates the ability of energy extraction
through parametric instability. The cross-flow can still
support transient growth, but the algebraic growth of perturbations is
expected to be overcome by the faster exponential growth provided by
the modal instability of~$U$. A total number of~100 uncorrelated flow
fields with modally unstable $U(y,z,t_0)$ were selected to run
simulations. Note that as the base flow is frozen in time, the
assumption of constant $U$ invoked for the stability analysis is
rigorously satisfied. As an example, Figure~\ref{fig:snap_modal} shows
the instantaneous velocity field for one case after transients.
%

The resulting root-mean-squared (rms) fluctuating velocities for the
statistical steady state are shown in Figure~\ref{fig:stats_modal}(a),
together with those from CH180.  Unsurprisingly, turbulent channel
flows with persistent modally unstable mean cross-flow are capable of
sustaining turbulence. The new flow reaches statistical equilibrium at
a higher level of turbulence intensities owing to the additional mean
tangential stress introduced by $f_1$, but the trends observed in
Figure~\ref{fig:stats_modal}(a) are consistent with CH180 in terms of
relative magnitude and wall-normal behavior. The transition to the new
steady state is evidenced by Figure~\ref{fig:stats_modal}(b), which
shows the time evolution of a selection of streamwise Fourier
components before and after freezing the mean cross-flow. The
adaptation time of turbulence upon imposition of constant $U$ is
roughly $h/u_\tau$, consistent with the lifespan of large eddies in
the flow \citep{Lozano2014b}.
%

The results reported above correspond to one particular $U(y,z,t_0)$,
but the conclusions are found to be robust for all mean cross-flows
examined.  Finally, it is important to highlight that while
maintaining mean cross-flow in a modally unstable state does lead to
sustained turbulence, whether this new state is similar in nature to
unforced wall turbulence is an important question that is not
investigated here and should be carefully addressed in future studies.

\subsection{Energy injection by transient growth}\label{subsec:numerical:nonmodal}

\begin{figure}
 \vspace{0.2cm}
 \begin{center}
   \includegraphics[width=1\textwidth]{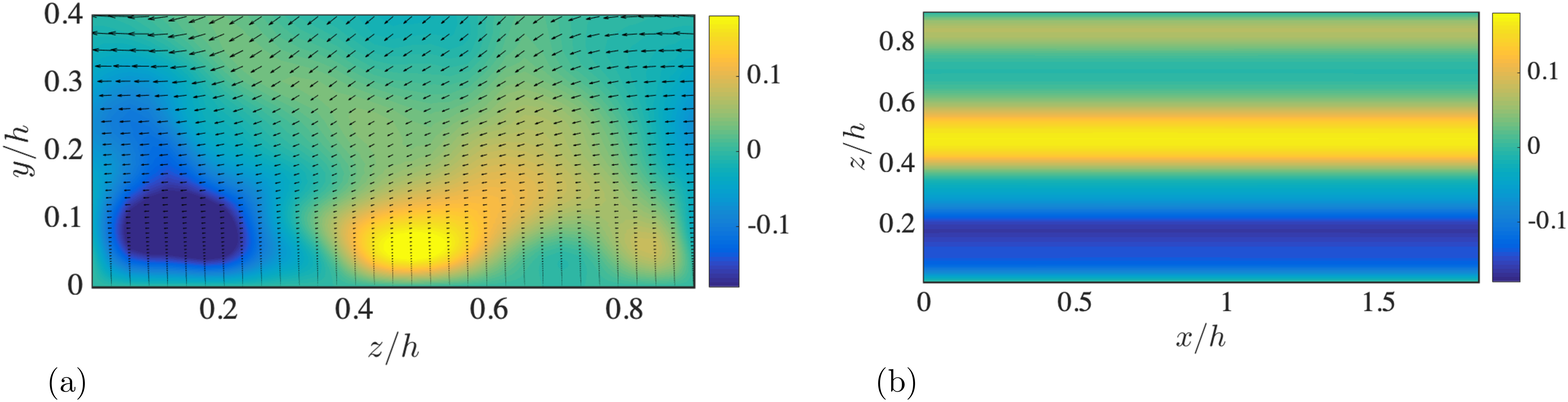}
 \end{center}
\caption{ Experiment with fixed, stable $U$: (a) Instantaneous
  velocity field in a $z-y$ plane at $x=0h$. (b) Instantaneous
  streamwise velocity in a $x-z$ plane at $y=0.1h$. Colors represent
  streamwise velocity and arrows are cross-flow velocities. Velocities
  are scaled in wall units of the baseline case. Arrows in panel (a)
  are amplified by a factor of 10.
\label{fig:snap_nonmodal}}
 \begin{center}
   \subfloat[]{ \includegraphics[width=0.45\textwidth]{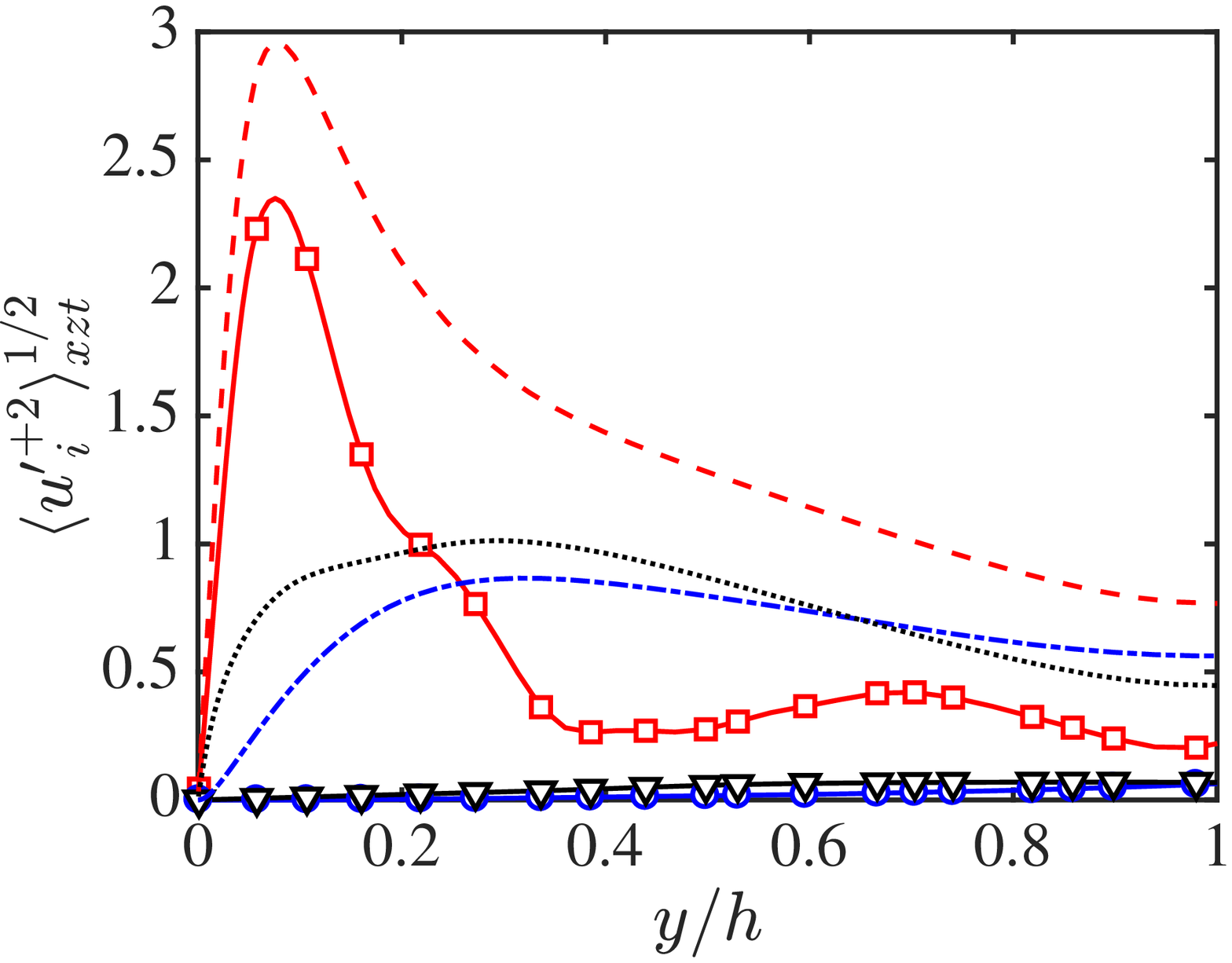} }
   \hspace{0.1cm}
   \subfloat[]{ \includegraphics[width=0.45\textwidth]{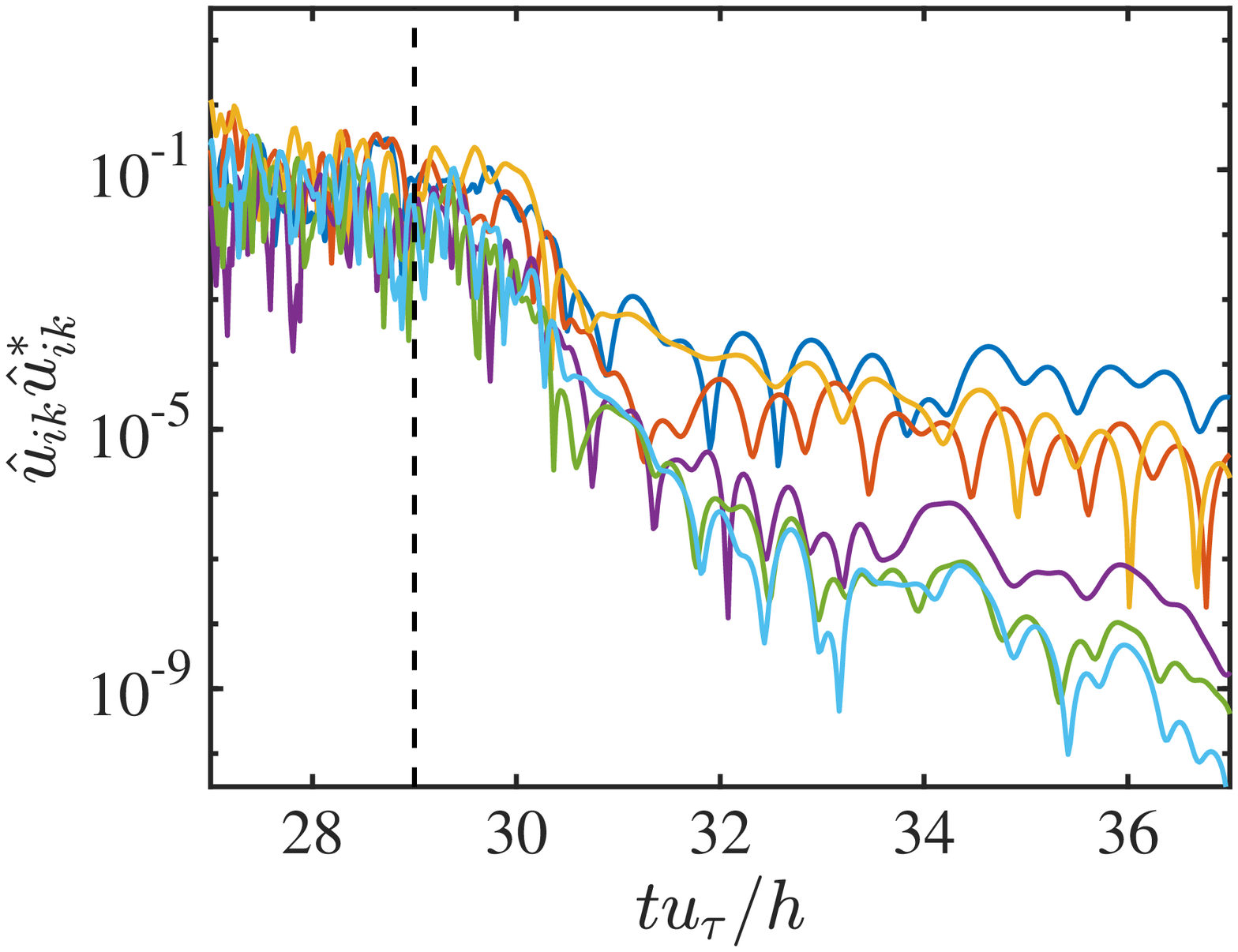} }
 \end{center}
\caption{ Experiment with fixed, stable $U$: (a) Root-mean-squared
  fluctuating velocities for case CH180 (lines) and channel with
  frozen-in-time modally stable mean cross-flow (symbols).  Lines and
  symbols are: dashed and squares, streamwise; dash-dotted and
  circles, wall-normal; dotted and triangles, spanwise velocity
  fluctuations.  (b) Time evolution of the energy associated to
  streamwise Fourier modes, $\hat u_{ik} \hat u^*_{ik}$, for $i=1,2,3$
  and $k=0,1,2$ at $y=0.1h$, where $*$ denotes complex conjugate. The
  mean cross-flow is frozen in time at $t u_\tau/h = 29$ (dashed black
  line).
\label{fig:stats_nonmodal}}
\end{figure}

The effect of non-modal transient growth as a main cause for energy
injection is assessed by following a similar approach to that in
Section~\ref{subsec:numerical:modal}. In this case, the cross-flow $U$
from CH180 is frozen at the instant~$t_0$, when the flow is modally
stable.  The mean flow $\langle u \rangle_{xzt}$ is set to the same
value as in case CH180. The set-up disposes of energy transfers that
are due to both modal and parametric instabilities, while maintaining
the transient growth of perturbations. The expected scenario
consistent with sustained turbulence \citep[e.g.,][]{Schoppa2002} is
the non-modal amplification of perturbations until saturation followed
by nonlinear scattering and generation of new disturbances. However,
plain visual inspection of the velocity field in
Figure~\ref{fig:snap_nonmodal} reveals that this is not the case, and
turbulence is distinctly lessened.

The rms fluctuating velocities for one experiment are shown in
Figure~\ref{fig:stats_nonmodal}(a).  Turbulence reaches a
quasi-laminar state with residual cross-flow turbulence intensities
and non-negligible streamwise fluctuations required to support the
prescribed $U(y,z,t_0)$. The exponential decay of Fourier modes after
freezing the mean cross-flow is clearly seen in Figure
\ref{fig:stats_nonmodal}(b).  The simulation was repeated for 20
different modally stable mean cross-flows $U(y,z,t_0)$ and all cases
decayed similarly to the example discussed above.
%

\subsection{Energy injection by transient growth with parametric instability}\label{subsec:numerical:parametric}

The maintenance of turbulence exclusively by transient growth with
parametric instability is analyzed by a time-dependent mean cross-flow
that is altered to be free of modal instabilities. To that end, we
introduce the linear damping $f_1=-\damp ( U - \langle
u\rangle_{xz})$, $f_2=f_3=0$, where the parameter $\damp$ is a
coefficient to be determined such that $U$ is modally stable for all
times.  The goal is to investigate the existence of self-sustained
wall turbulence without any energy extraction from the mean cross-flow
via modal instabilities.

Ideally, if $\partial \boldsymbol{u}' \big/ \partial t = A(\damp)
\boldsymbol{u}'$ is the linear equation governing the fluctuating
velocities, the drag coefficient~$\damp$ should be adjusted at each
time step to bring the most unstable eigenvalue of $A$ to neutrality.
In the present preliminary version of the work, we adopted a simplified
approach where the value of $\damp$ is set constant in time. Then, a
campaign of channel flow simulations driven by a constant streamwise
mass flux was performed for values of $\damp$ ranging from $0$ up to
$\damp_c \approx 1.3 u_\tau/h$, above which the flow laminarizes. 
%
\begin{figure}
 \begin{center}
   \subfloat[]{ \includegraphics[width=0.31\textwidth]{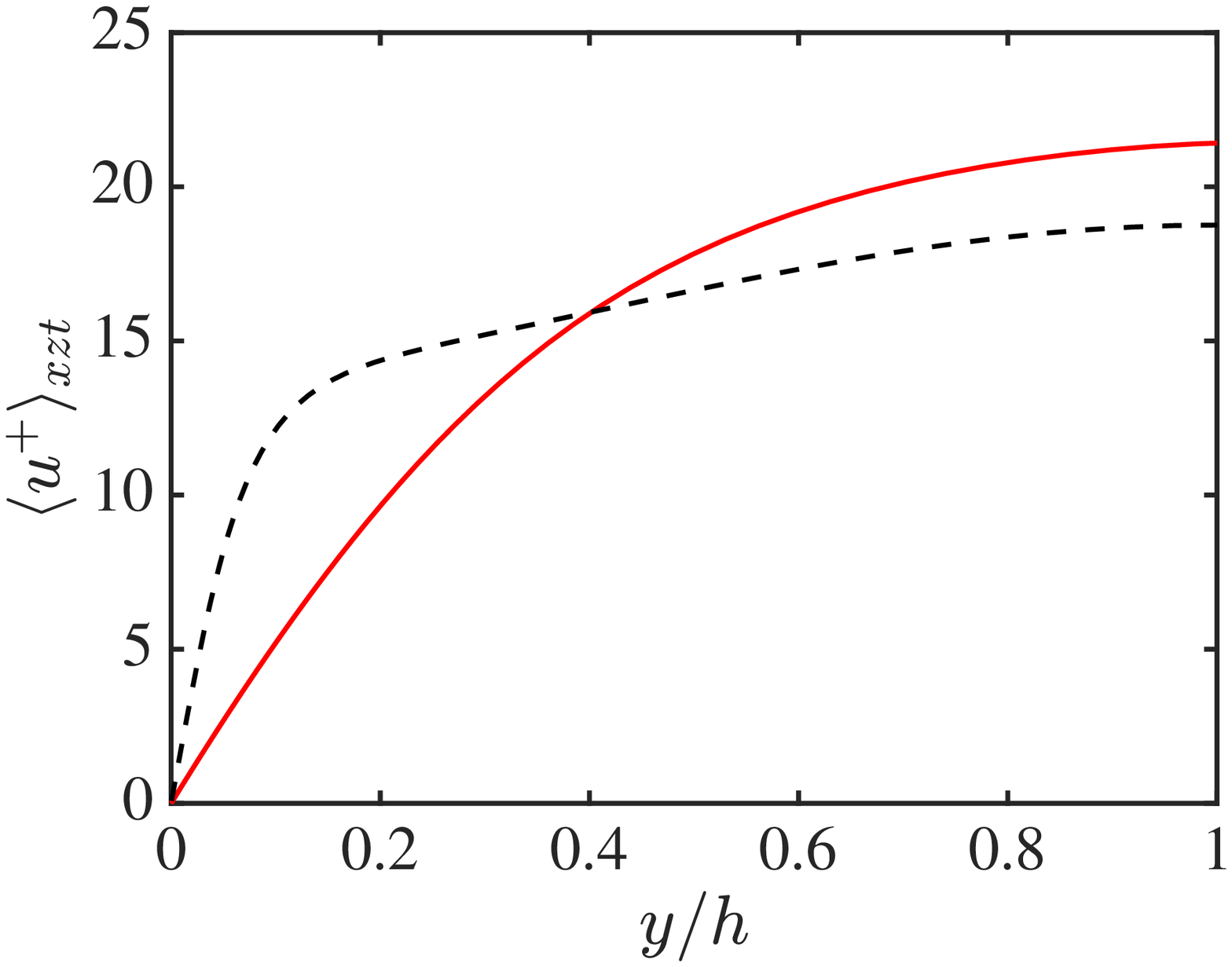} }
   \hspace{0.05cm}
   \subfloat[]{ \includegraphics[width=0.31\textwidth]{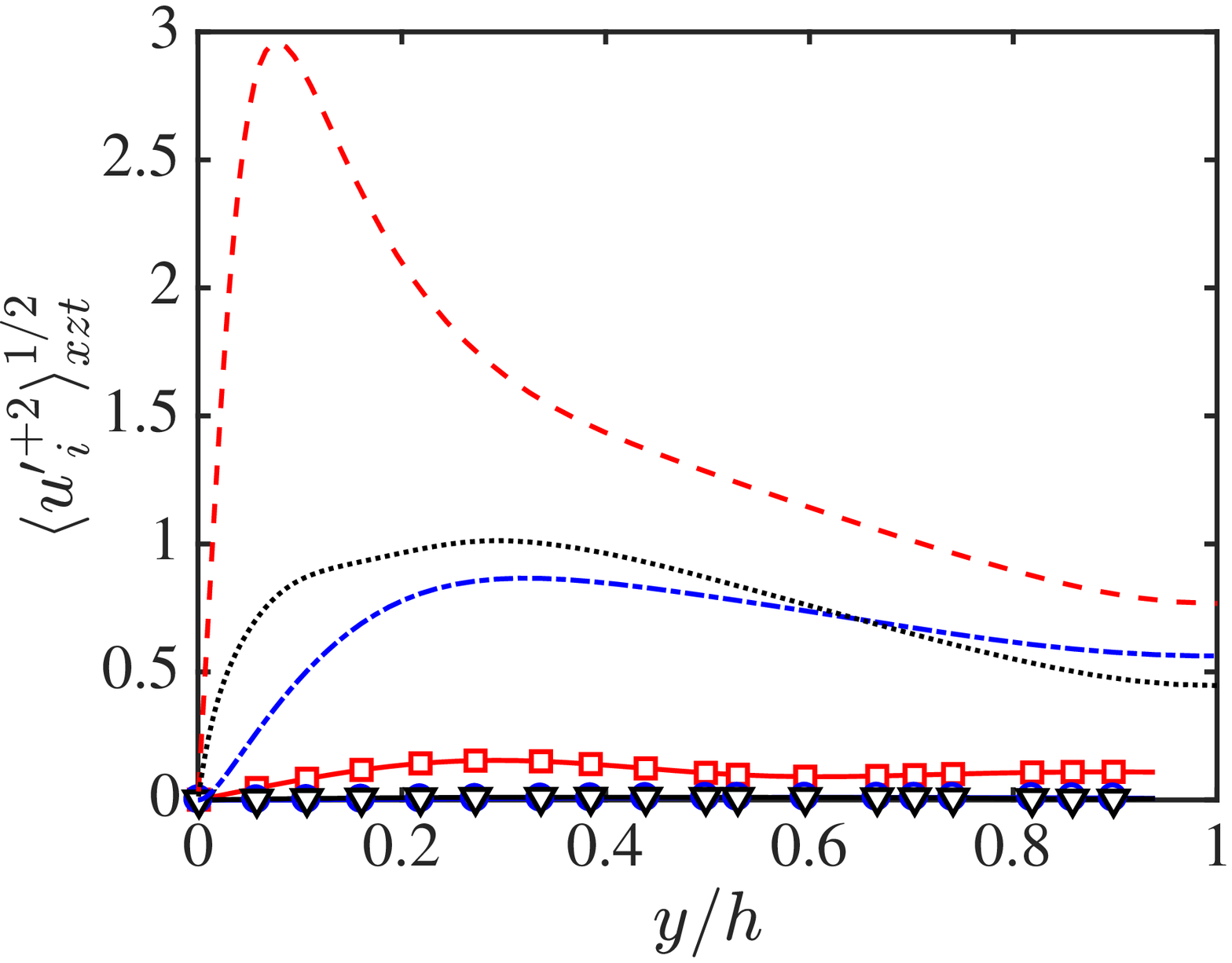} }
   \hspace{0.05cm}
   \subfloat[]{ \includegraphics[width=0.31\textwidth]{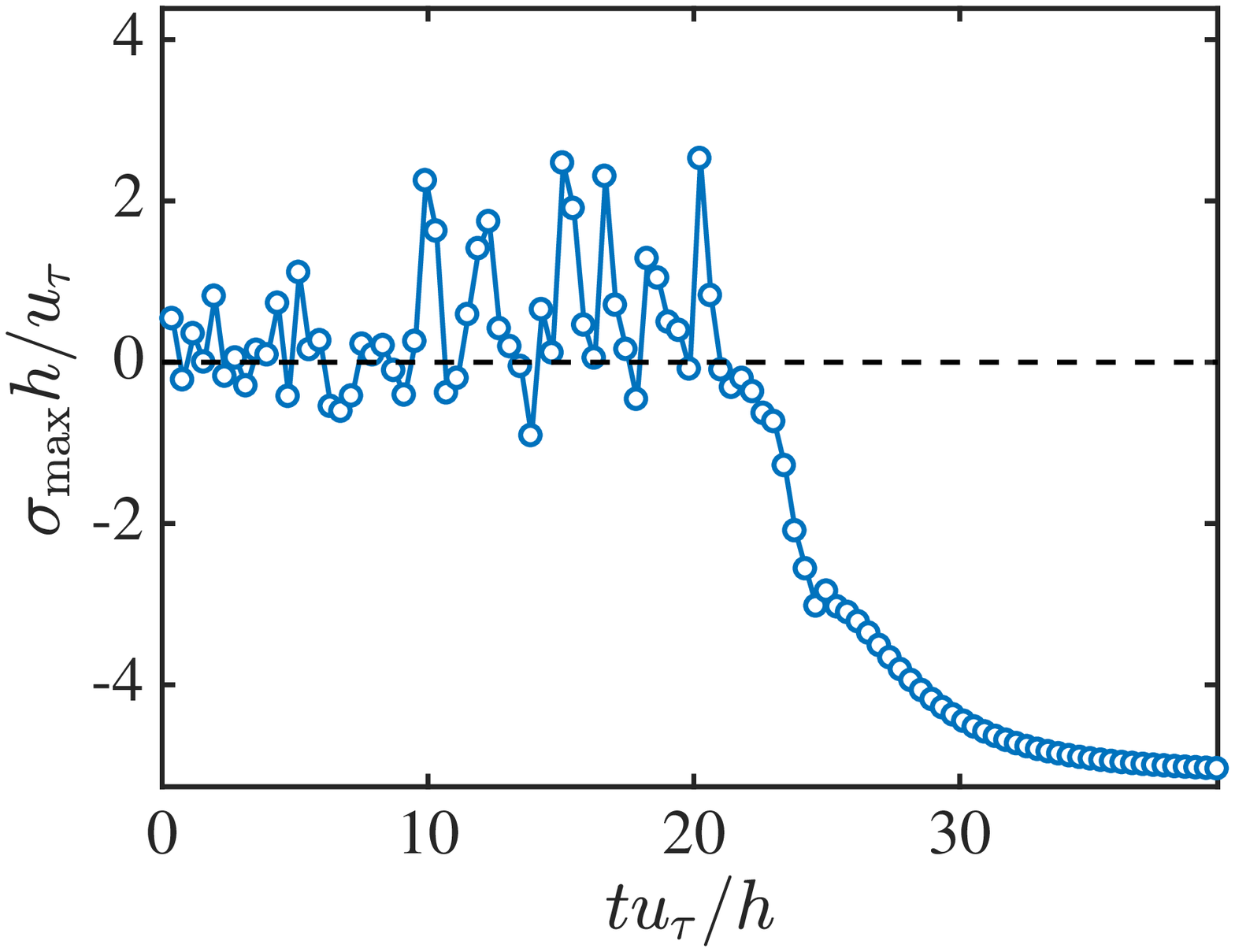} }
 \end{center}
 \begin{center}
   \subfloat[]{ \includegraphics[width=0.31\textwidth]{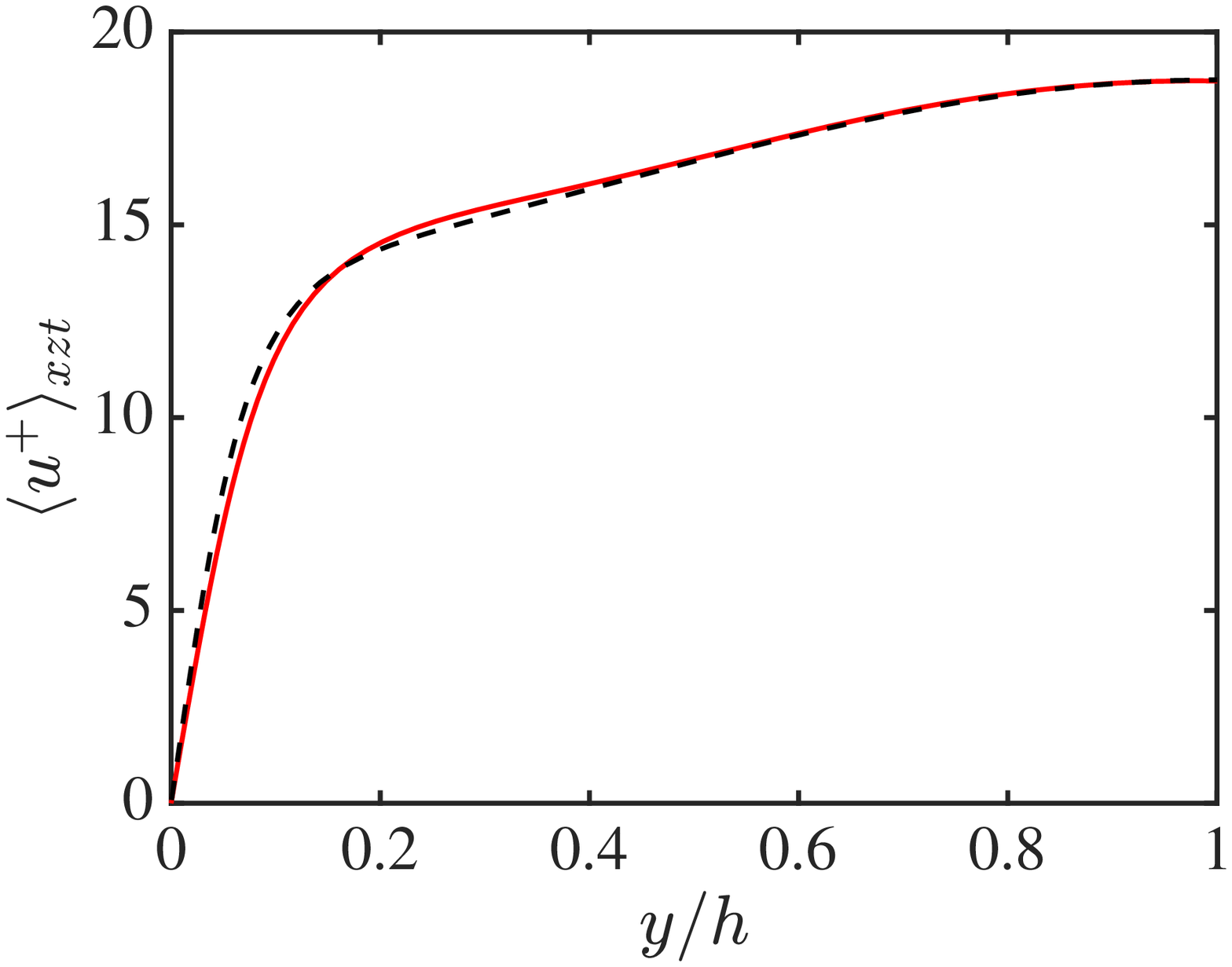} }
   \hspace{0.05cm}
   \subfloat[]{ \includegraphics[width=0.31\textwidth]{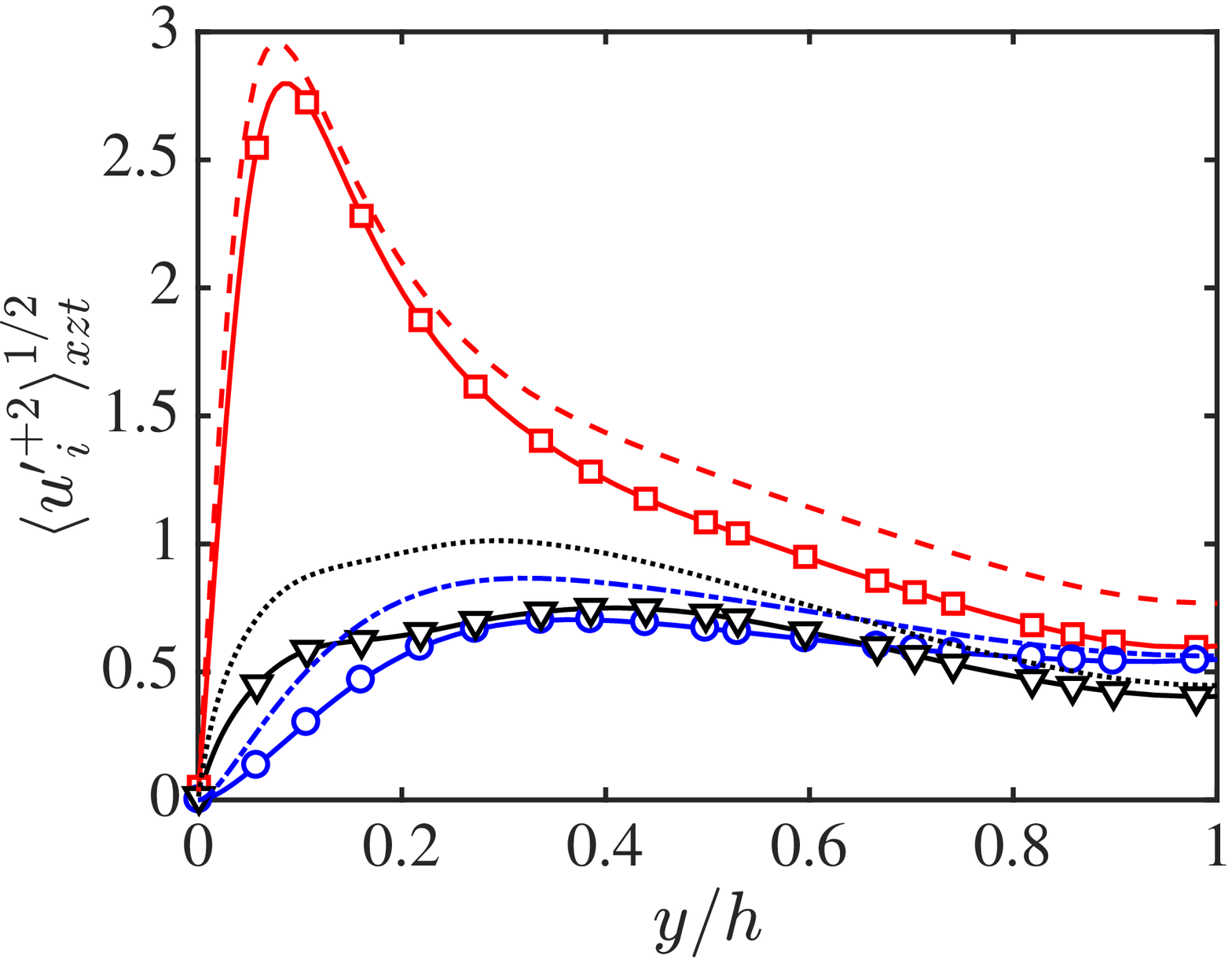} }
   \hspace{0.05cm}
   \subfloat[]{ \includegraphics[width=0.31\textwidth]{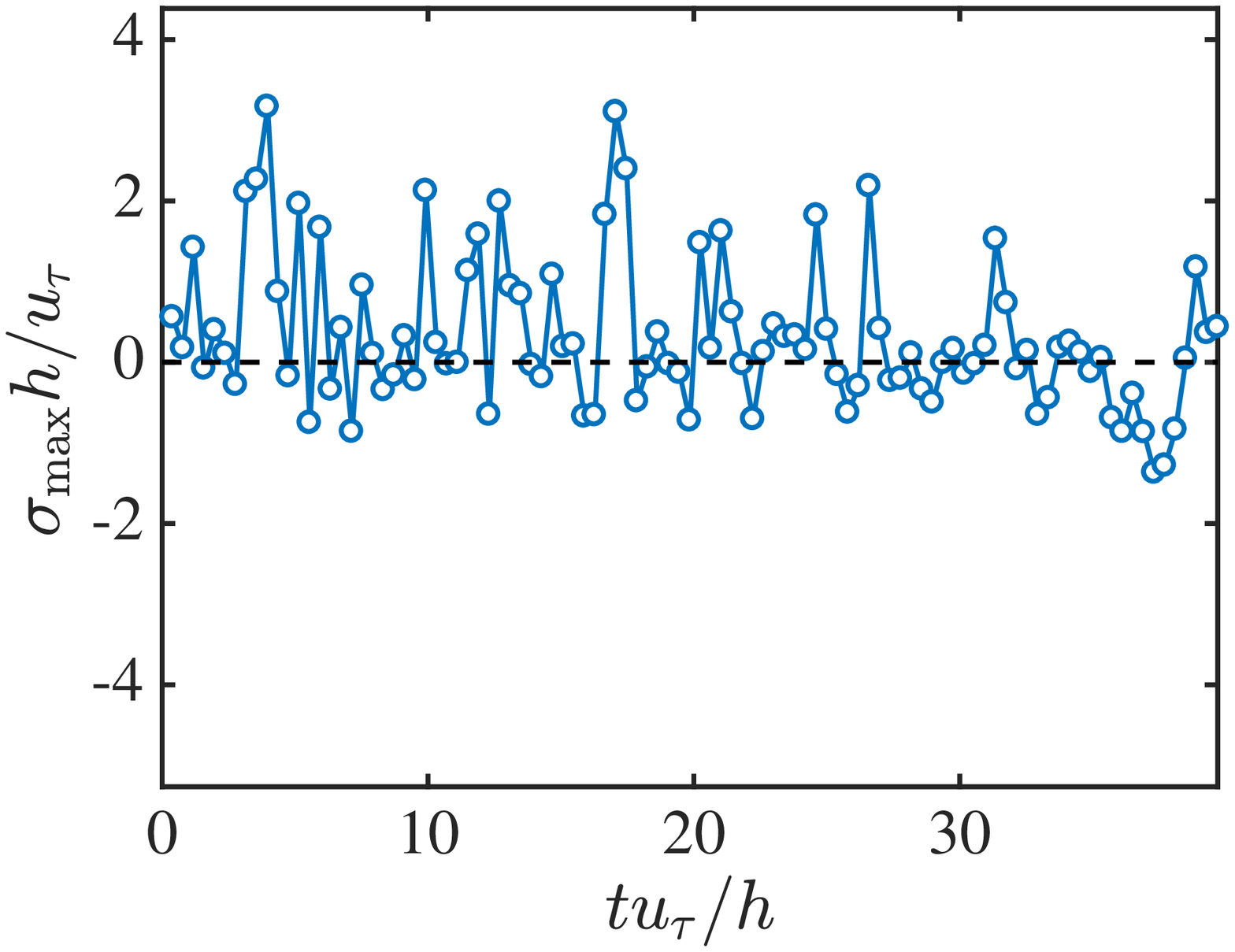} }
 \end{center}
%
\caption{ Experiment with linear drag on $U$: (a,d) Mean velocity
  profile for case CH180 (black dashed line) and channel with linear
  damping $-\damp (U - \langle u\rangle_{xz})$ (solid red line). (b,e)
  Root-mean-squared fluctuating velocities for case CH180 (lines) and
  channel with $-\damp (U - \langle u\rangle_{xz})$ (symbols). Lines
  and symbols are: dashed and squares, streamwise; dashed-dotted and
  circles, wall-normal; dotted and triangles, spanwise velocity
  fluctuations. (c,g) Time evolution of the maximum growth rate of $A$
  for channel flow with linear damping $-\damp (U - \langle
  u\rangle_{xz})$. (a,b,c) are for $\damp=1.4 u_\tau/h>\damp_c$ and
  (d,e,f) are for $\damp=1.2 u_\tau/h<\damp_c$.
\label{fig:stats_parametric_1}}
\end{figure}

The mean and rms velocity profiles for $\damp=1.4 u_\tau/h>\damp_c$
are shown in Figures~\ref{fig:stats_parametric_1}(a,b).  The flow is
laminar with zero velocity fluctuations.
Figure~\ref{fig:stats_parametric_1}(c) shows the time history of the
most unstable growth rate of $A$, which is constant and negative after
transients.  Figures~\ref{fig:stats_parametric_1}(d,e,f) are
equivalent to Figures~\ref{fig:stats_parametric_1}(a,b,c) but for
$\damp=1.2 u_\tau/h<\damp_c$, which is the maximum value of $\damp$
that allows for sustained turbulence in a statistical steady state.
The rms velocities are weaker with respect to case CH180, but they
still resemble qualitatively those encountered in real
turbulence. Although not shown, the de-correlation times for $U$ are
similar to those for case CH180.
Figure~\ref{fig:stats_parametric_1}(f) shows that $U(z,y,t)$ is
modally unstable $\sim$60\% of the time based on
$\sigma_{\mathrm{max}}>u_\tau/h/4$. The percentage is below the value
obtained for case CH180 ($\sim$80\%), which suggests that not all the
modal instabilities are necessary to maintain turbulence with
realistic one-point statistics.

\begin{figure}
\begin{center}
  \subfloat[]{ \includegraphics[width=0.31\textwidth]{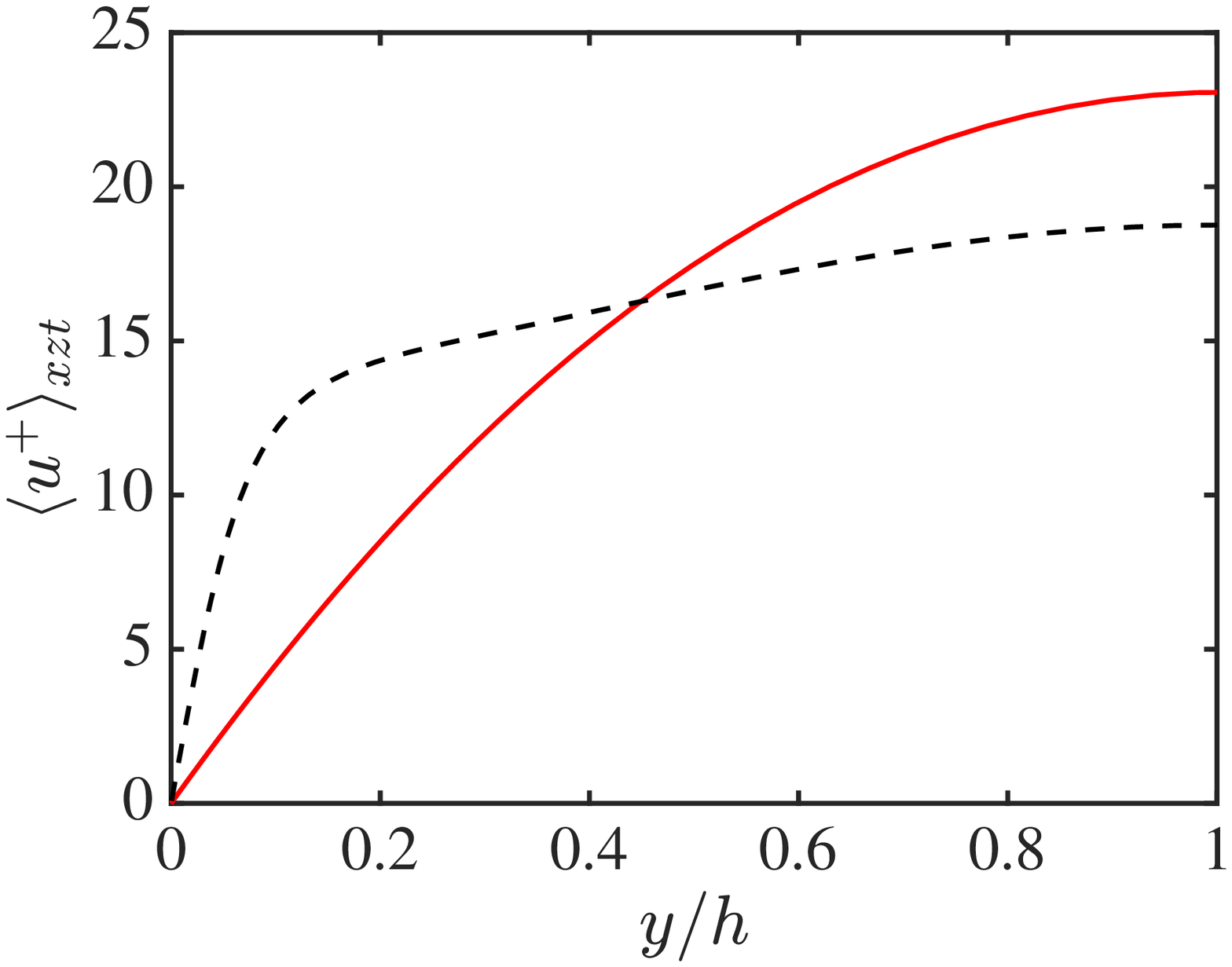} }
  \hspace{0.05cm}
  \subfloat[]{ \includegraphics[width=0.31\textwidth]{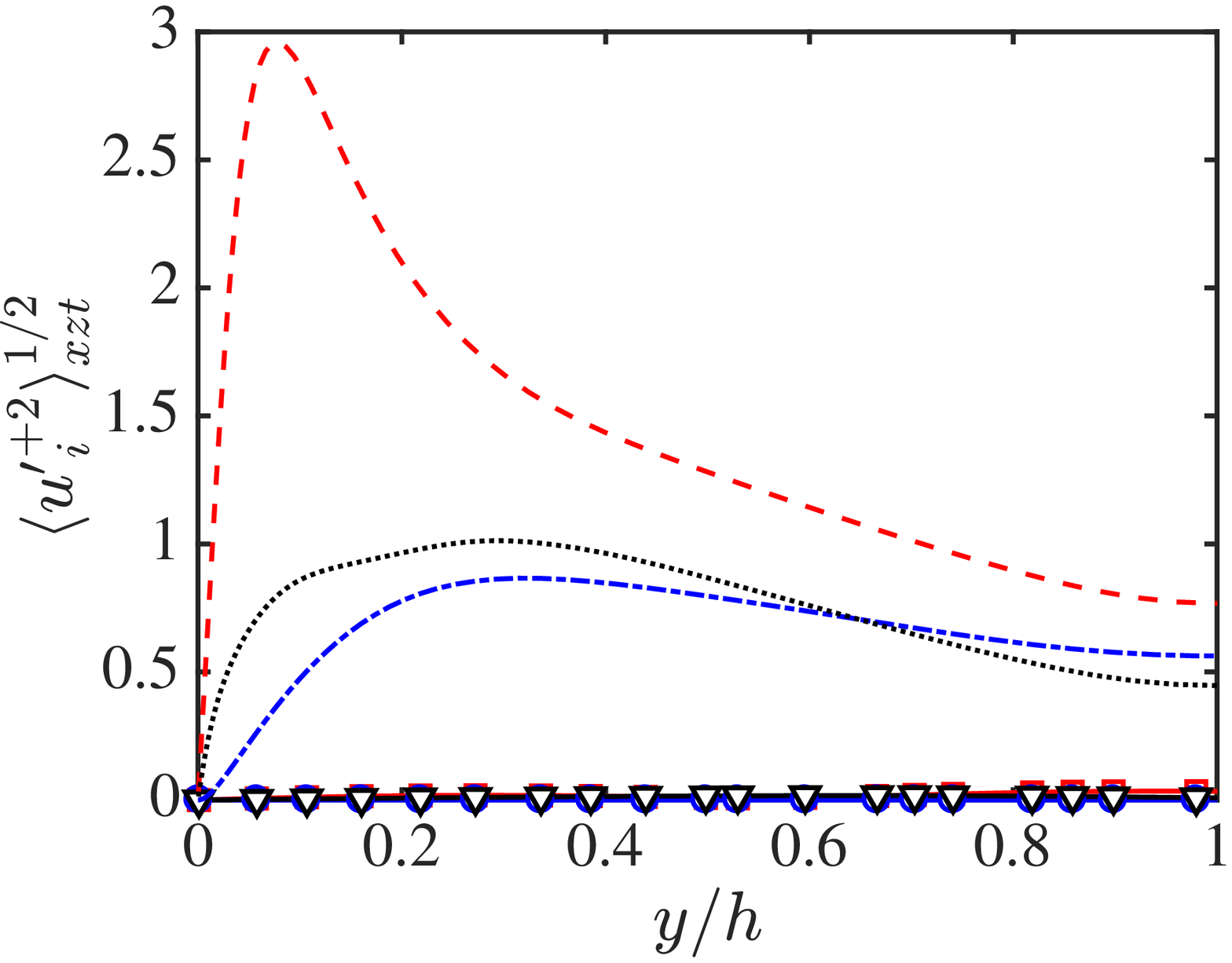} }
  \hspace{0.05cm}
  \subfloat[]{ \includegraphics[width=0.31\textwidth]{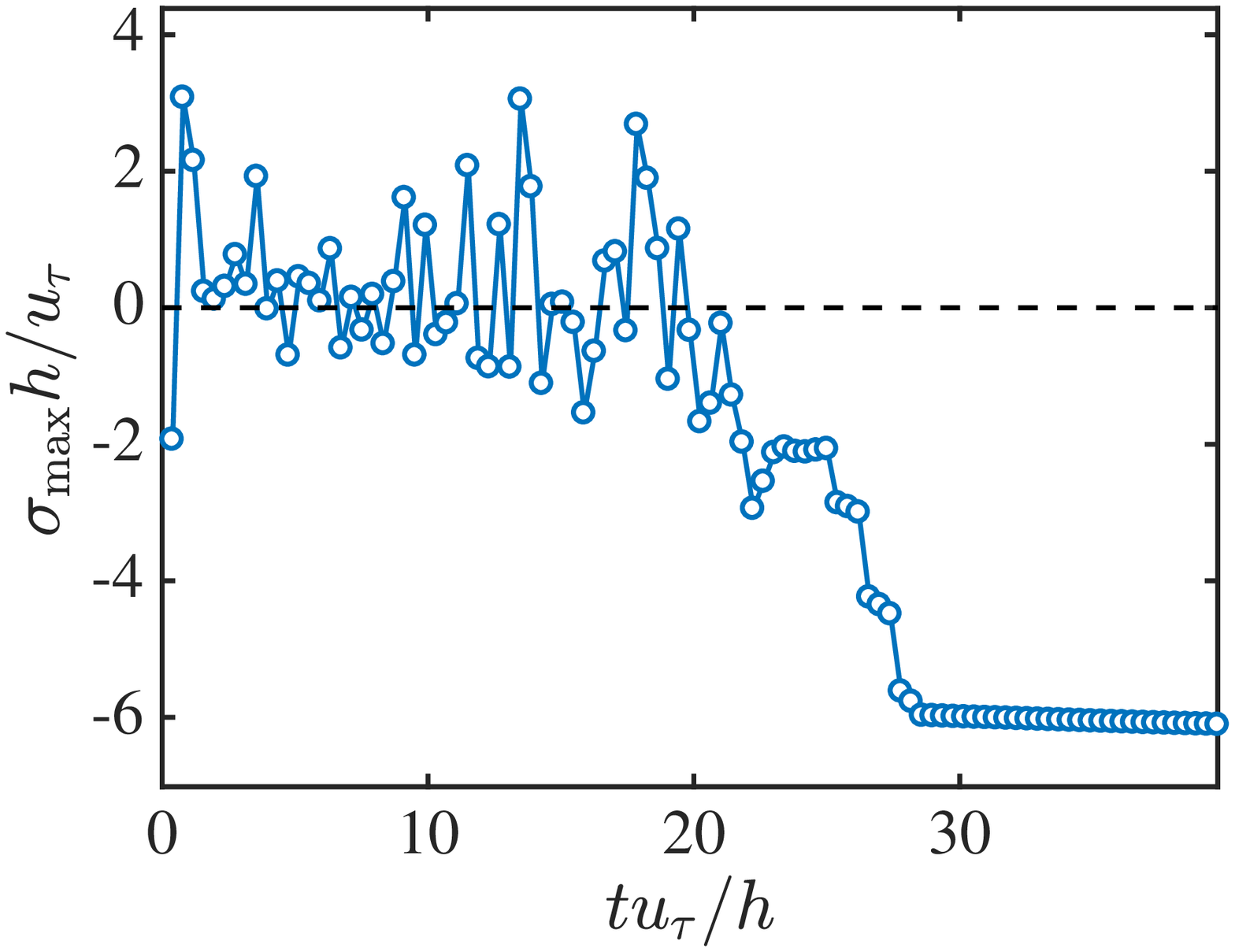} }
\end{center}

\begin{center}
  \subfloat[]{ \includegraphics[width=0.31\textwidth]{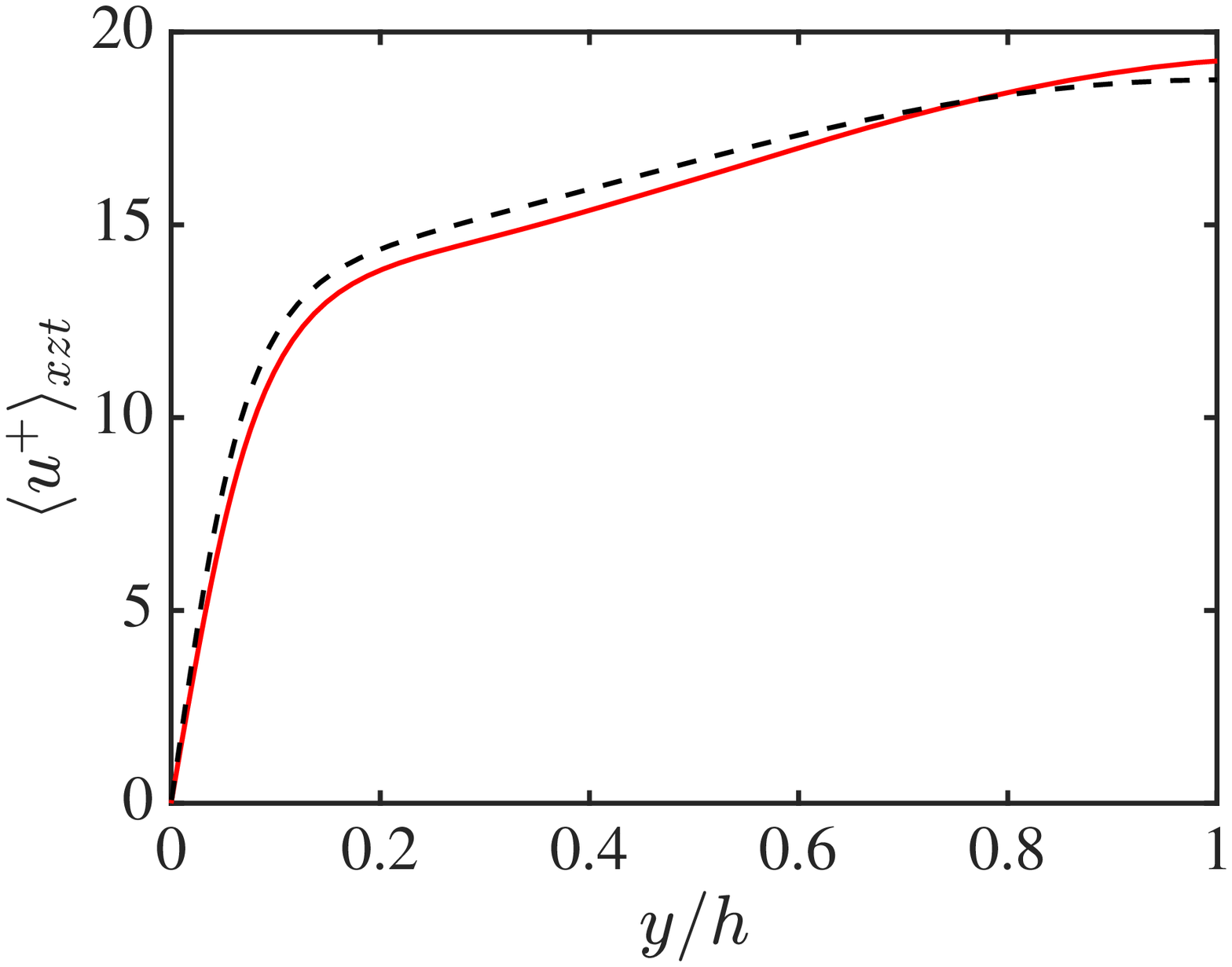} }
  \hspace{0.05cm}
  \subfloat[]{ \includegraphics[width=0.31\textwidth]{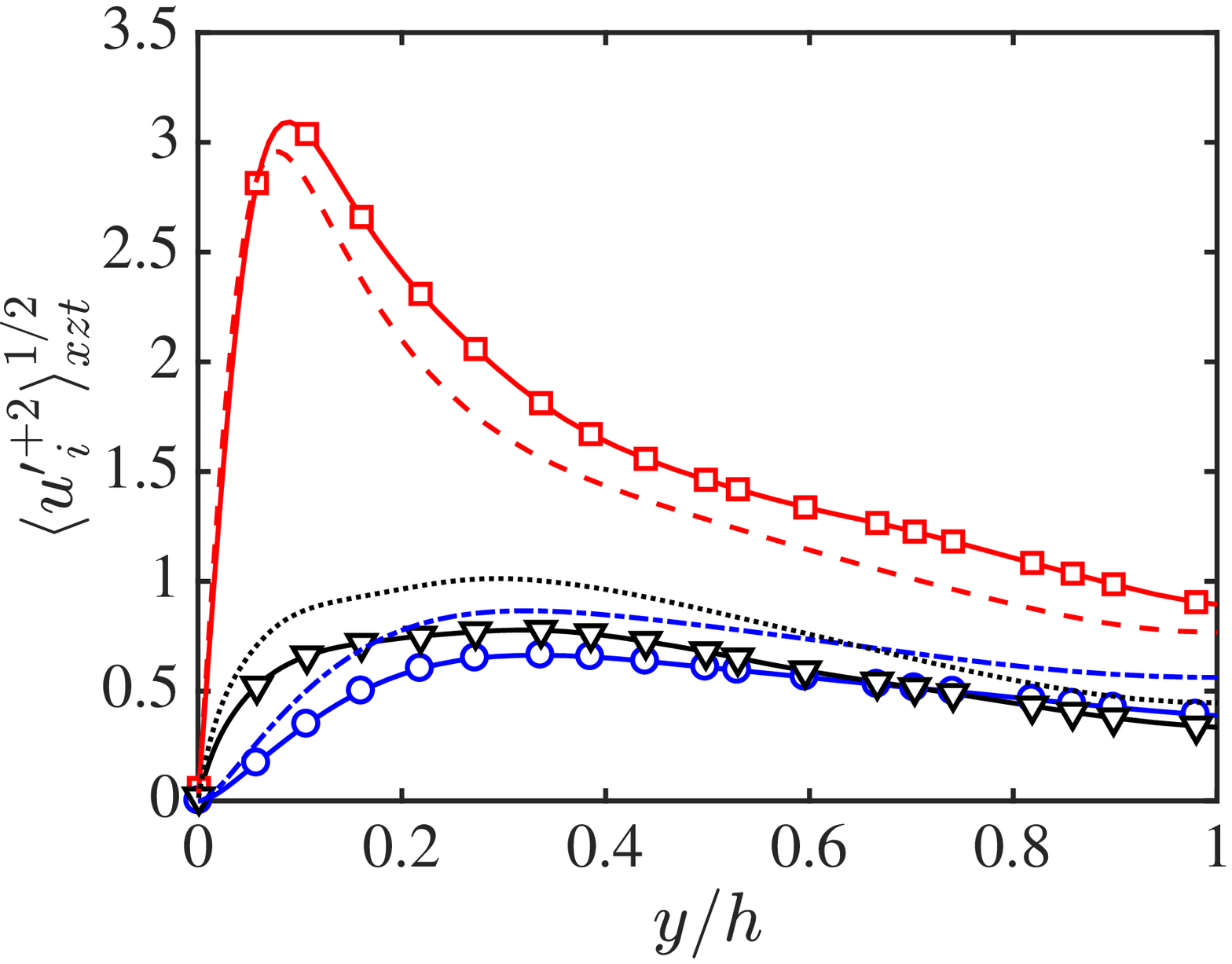} }
  \hspace{0.05cm}
  \subfloat[]{ \includegraphics[width=0.31\textwidth]{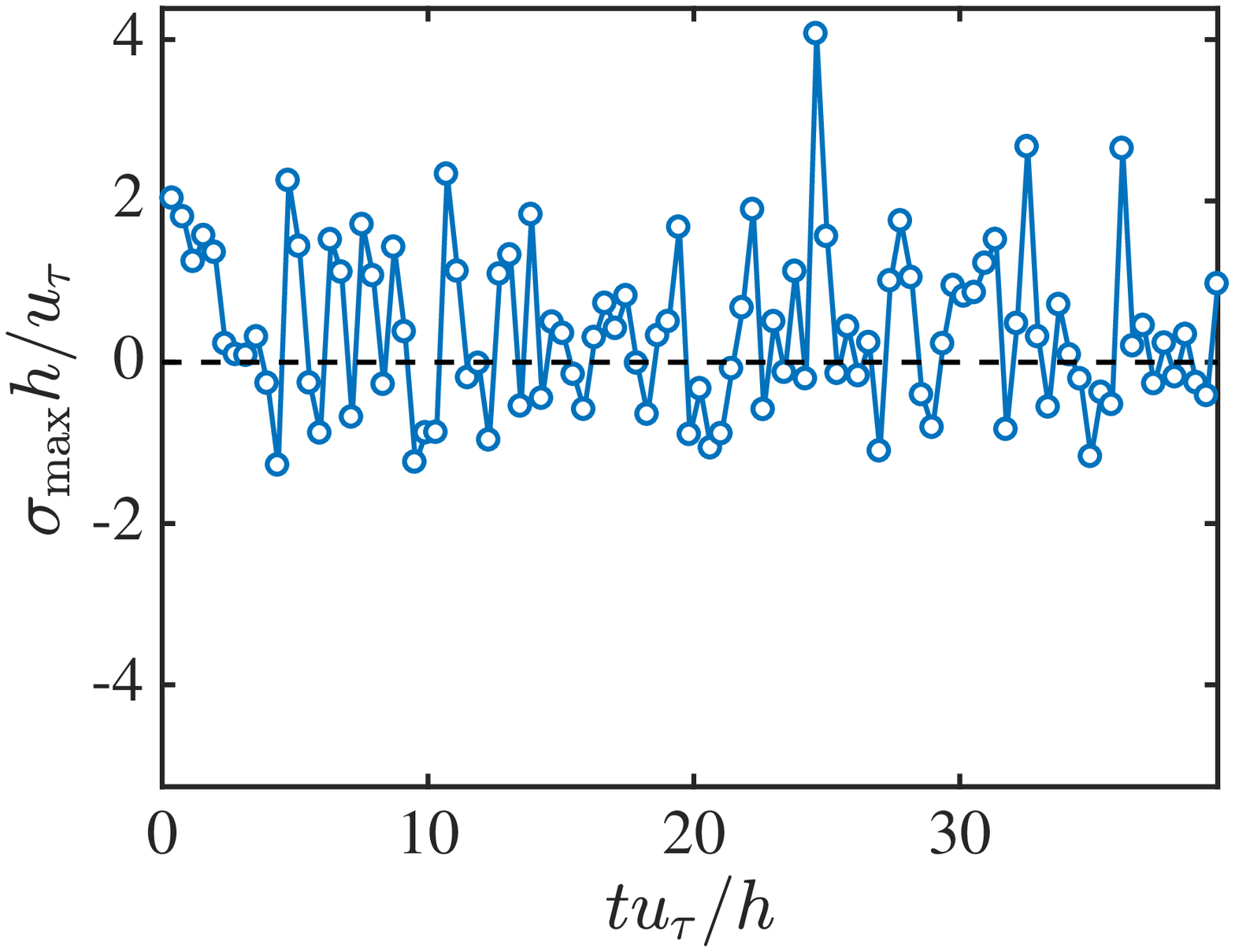} }
\end{center}
%
\caption{ Experiment with linear drag on $\boldsymbol{u}'$: (a,d) Mean
  velocity profile for case CH180 (black dashed line) and channel with
  linear damping $-\damp' u'_i$ (solid red line). (b,e)
  Root-mean-squared fluctuating velocities for case CH180 (dashed
  lines) and channel with $-\damp' u'_i$ (symbols).  Lines and symbols
  are: dashed and squares, streamwise; dashed-dotted and circles,
  wall-normal; dotted and triangles, spanwise velocity
  fluctuations. (c,g) Time evolution of the maximum growth rate of $A$
  for channel flow with linear damping $-\damp' u'_i$. (a,b,c) are for
  $\damp'=1.1 u_\tau/h>\damp_c'$ and (d,e,f) are for $\damp'= 0.9
  u_\tau/h<\damp_c'$.
\label{fig:stats_parametric_2}}
\end{figure}

Finally, a different numerical experiment is performed by including a
linear damping into the equation for the fluctuating velocities,
i.e.,~$f_i=-\damp' u_i'$, $i=1,2,3$. In this new set-up, we directly
target the eigenvalues of $A$, whose real parts are reduced exactly by
$\damp'$ compared to the eigenvalues of $A$ for CH180. The maximum
value of $\damp'$ that allows for sustained turbulence is found to be
$\damp'_c \approx 1 u_\tau/h$. The resulting flow statistics for
$\damp'$ that is marginally above and marginally below $\damp'_c$
(Figure \ref{fig:stats_parametric_2}) yield similar conclusions as
those reported above: turbulence only survives when $A$ is modally
unstable (based on $\sigma_{\mathrm{max}}>u_\tau/h/4$) for a
substantial fraction of the time simulated, in this case for
$\sim$50\% of the time when $\damp' = 0.9 h/u_\tau < \damp'_c$.
%

\section{Conclusions}\label{sec:conclusions}

We have studied the mechanism of energy injection from the mean flow
to the fluctuating velocity necessary to maintain wall
turbulence. This process is believed to be correctly represented by
the linearized Navier--Stokes equations, and three potential linear
mechanisms have been considered, namely, modal instability of the
streamwise mean cross-flow $U(y,z,t)$, non-modal transient growth, and
non-modal transient growth supported by parametric instability.

We have designed three numerical experiments of plane turbulent
channel flow with additional forcing terms aiming to neutralize one or
various linear mechanisms for energy extraction. To assess the effect
of modal instabilities and non-modal transient growth of $U(y,z,t)$,
we have computed turbulent channel flows with prescribed modally
stable/unstable mean cross-flows frozen in time. In addition,
transient growth with parametric instability was evaluated by adding a
linear damping to the momentum equation of the mean cross-flow or to
the fluctuation equations. This additional linear damping was chosen
accordingly to render any modal instabilities stable and thus preclude
energy transfer to the fluctuations from modal instabilities.

From our preliminary experiments, only cases with mean cross-flows
capable of supporting modal instabilities were found to sustain
turbulence. However, the question whether such a new turbulence
complies with the same physical mechanisms as those occurring in
actual (unforced) turbulence remains unanswered. On the other hand,
cases exclusively supported by transient growth decayed until
laminarization. For this preliminary study, this outcome should not be
taken as a demonstration that transient growth alone aided or not by
parametric instability is unable to maintain turbulence in actual
flows, but just as an indication that we could not find a
self-sustained turbulent system without the contribution of modal
instabilities.

Future work will be devoted to the careful design of modified
turbulent channel flows providing clear causal inference and
quantification of the energy injection mechanisms in wall
turbulence. Moreover, if indeed modal instability (or other) is the
dominant mechanism responsible for transferring energy from the mean
flow to the fluctuations, it should be detectable from unforced
wall-turbulence simulation (e.g., CH180), and additional efforts will
be carried on to analyze DNS data using non-intrusive techniques. 



\section*{Acknowledgments}

The work was supported by NASA under Grant~\#NNX15AU93A and ONR under
Grant~\#N00014-16-S-BA10.

\section*{Appendix: Stability analysis}

\vspace{0.1in} \setcounter{saveequation}{\value{equation}}
\stepcounter{saveequation}\setcounter{equation}{0}
\renewcommand{\theequation}{A \arabic{equation}}

This appendix describes the linear stability analysis of a base
mean cross-flow, which is inhomogeneous in two spatial directions. We assume
the following velocity field
\beq \bu=\left(U(y,z),0,0\right)+\e\bu_d, \eeq where the base flow
$U$ is assumed parallel, steady, and streamwise independent, and
$\bu_d$ is the disturbance. Substituting the velocity field into
the incompressible Navier--Stokes equations and neglecting the
nonlinear terms, we obtain
\bse\label{ns}
\begin{align}
\frac{\p u_d}{\p x}+\frac{\p v_d}{\p y}+\frac{\p w_d}{\p
z} & =0, \\
\frac{\p u_d}{\p t} + U\frac{\p u_d}{\p x}+v_d\frac{\p U}{\p
y} + w_d\frac{\p U}{\p z} &= -\frac{\p p_d}{\p
x} + \nu\nabla^2 u_d, \\
\frac{\p v_d}{\p t} + U\frac{\p v_d}{\p x} &= -\frac{\p p_d}{\p
y} + \nu\nabla^2 v_d, \\
\frac{\p w_d}{\p t} + U\frac{\p w_d}{\p x} &= -\frac{\p p_d}{\p
z} + \nu\nabla^2 w_d, \end{align}
\ese
where $p_d$ is the disturbance pressure. The boundary conditions are no slip and
impermeability on the channel walls. In the current study, the
stability analysis has been performed only on a half-channel.
Therefore, no slip and impermeability were imposed on the channel
center, as we are interested only in the instabilities close to
the wall.

The base flow is periodic along the spanwise direction, and it is
often useful to describe it in terms of a truncated Fourier
expansion. In such cases, a Floquet analysis is performed with respect
to the span \citep[see, e.g.,][]{Karp2014}. Nevertheless, for an
arbitrary base flow, such as the one considered here, it is not
beneficial to invoke Floquet theory. Therefore, we assume the
following form for the disturbance,
\beq \bq_d=\hat{\bq}_d(y,z)e^{\lambda t + i\alpha x}, \eeq
where $\bq_d=(u_d,v_d,w_d,p_d)^\mathsf{T}$, $\alpha$ is the streamwise
wavenumber, and $\lambda$ is the temporal complex eigenvalue. The
eigenvalue can be written as $\lambda=\sigma+i\omega$, where $\sigma$
is the growth rate and $\omega$ is the frequency. The linearized
equations above are discretized along both inhomogeneous directions
using spectral methods. Along the wall-normal direction, a Chebyshev
grid is used for $y\in[0,h]$, and along the spanwise direction a
Fourier grid is used for $z\in[0,L_z]$.

Substituting the disturbance into the linearized equations, they
can be rearranged as a generalized eigenvalue problem for the calculation of
$\lambda$,
\beq \begin{pmatrix}
\mathsfbi{D_x}~&\mathsfbi{D_y}~&\mathsfbi{D_z}~&\mathsfbi{O}\\
\mathsfbi{C}~&\mathsfbi{U_y}~&\mathsfbi{U_z}~&\mathsfbi{D_x}\\
\mathsfbi{O}~&\mathsfbi{C}~&\mathsfbi{O}~&\mathsfbi{D_y}\\
\mathsfbi{O}~&\mathsfbi{O}~&\mathsfbi{C}~&\mathsfbi{D_z}
\end{pmatrix} \begin{pmatrix}
\ba{c}\tilde{u}_d\\\tilde{v}_d\\\tilde{w}_d\\\tilde{p}_d
\ea \end{pmatrix} = \lambda \begin{pmatrix}\mathsfbi{O}~&\mathsfbi{O}~&\mathsfbi{O}~&\mathsfbi{O}\\
-\mathsfbi{I}~&\mathsfbi{O}~&\mathsfbi{O}~&\mathsfbi{O}\\
\mathsfbi{O}~&-\mathsfbi{I}~&\mathsfbi{O}~&\mathsfbi{O}\\
\mathsfbi{O}~&\mathsfbi{O}~&-\mathsfbi{I}~&\mathsfbi{O} \end{pmatrix} 
\begin{pmatrix}\tilde{u}_d\\\tilde{v}_d\\\tilde{w}_d\\\tilde{p}_d
\end{pmatrix}. \eeq Here, $\mathsfbi{I}$ is the identity matrix,
$\mathsfbi{O}$ is a zero matrix, $\tilde{u}_d$ (and similarly
$\tilde{v}_d,\tilde{w}_d,\tilde{p}_d$) is a one-dimensional
representation of a two-dimensional vector
\beq
\tilde{u}_d=\big(\hat{u}_d(y,z_1),\hat{u}_d(y,z_2),\dots,\hat{u}_d(y,z_{N_z})\big)^\mathsf{T},
\eeq
and the matrices $\mathsfbi{C}$, $\mathsfbi{U_y}$,
$\mathsfbi{U_z}$, $\mathsfbi{D_x}$, $\mathsfbi{D_y}$, and
$\mathsfbi{D_z}$ are given by
\bse\begin{align}
\mathsfbi{C} &= i\alpha\;\textnormal{diag}\left(\mathsfbi{U}\right) -
\nu\left(\mathsfbi{\bar{I}_z}\otimes\mathsfbi{\bar{D}^2_y}
+ \mathsfbi{\bar{D}^2_z}\otimes\mathsfbi{\bar{I}_y} -
\alpha^2\mathsfbi{\bar{I}_z}\otimes\mathsfbi{\bar{I}_y}\right),\\
\mathsfbi{U_y} &=\textnormal{diag}\left\{\left(
\mathsfbi{\bar{I}_z}\otimes\mathsfbi{\bar{D}_y}\right)\mathsfbi{U}\right\},\\
\mathsfbi{U_z} &=\textnormal{diag}\left\{\left(
\mathsfbi{\bar{D}_z}\otimes\mathsfbi{\bar{I}_y}\right)\mathsfbi{U}\right\},\\
\mathsfbi{D_x} &=i\alpha\;\mathsfbi{\bar{I}_z}\otimes\mathsfbi{\bar{I}_y},\\
\mathsfbi{D_y} &=\mathsfbi{\bar{I}_z}\otimes\mathsfbi{\bar{D}_y},\\
\mathsfbi{D_z} &=\mathsfbi{\bar{D}_z}\otimes\mathsfbi{\bar{I}_y},
\end{align} \ese
where $\otimes$ is the Kronecker product and $\mathsfbi{U}$ is a
one-dimensional representation of $U$ (similarly to
$\tilde{u}_d$). The matrices $\mathsfbi{\bar{I}_y}$ and
$\mathsfbi{\bar{I}_z}$ are identity matrices of dimensions $N_y\times
N_y$ and $N_z\times N_z$, respectively, and $\mathsfbi{\bar{D}_y}$ and
$\mathsfbi{\bar{D}_z}$ are matrices that represent derivation with
respect to the $y$ and $z$ coordinates, respectively. The eigenvalue
problem is solved numerically using the software \textsc{Matlab}, with
$N_y=101$ and $N_z=32$. All the calculations were conducted for
$\alpha = 2\pi/L_x$.


\bibliographystyle{ctr}

\begin{thebibliography}{0}
\expandafter\ifx\csname natexlab\endcsname\relax\def\natexlab#1{#1}\fi

\end{thebibliography}


\begin{thebibliography}{38}
\expandafter\ifx\csname natexlab\endcsname\relax\def\natexlab#1{#1}\fi

\bibitem[Adrian(2007)]{Adrian2007}
{\sc Adrian, R.~J.} 2007 Hairpin vortex organization in wall turbulence. {\em
  Phys. Fluids\/} {\bf 19}, 041301.

\bibitem[del \'Alamo \& Jim{\'e}nez(2006)]{DelAlamo2006}
{\sc del \'Alamo, J.~C. \& Jim{\'e}nez, J.} 2006 Linear energy amplification in
  turbulent channels. {\em J. Fluid Mech.\/} {\bf 559}, 205--213.

\bibitem[Bae {\em et~al.\/}(2018{\natexlab{{\em a\/}}})Bae, Lozano-Dur\'an,
  Bose \& Moin]{Bae2018b}
{\sc Bae, H.~J., Lozano-Dur\'an, A., Bose, S.~T. \& Moin, P.}
  2018{\natexlab{{\em a\/}}} Dynamic slip wall model for large-eddy simulation.
  {\em J. Fluid Mech.\/} {\bf 859}, 400--432.

\bibitem[Bae {\em et~al.\/}(2018{\natexlab{{\em b\/}}})Bae, Lozano-Dur\'an,
  Bose \& Moin]{Bae2018}
{\sc Bae, H.~J., Lozano-Dur\'an, A., Bose, S.~T. \& Moin, P.}
  2018{\natexlab{{\em b\/}}} Turbulence intensities in large-eddy simulation of
  wall-bounded flows. {\em Phys. Rev. Fluids\/} {\bf 3}, 014610.

\bibitem[Butler \& Farrell(1992)]{Butler1992}
{\sc Butler, K.~M. \& Farrell, B.~F.} 1992 Optimal perturbations and streak
  spacing in wall-bounded turbulent shear flow. {\em Phys. Fluids A\/} {\bf 5},
  774.

\bibitem[Chernyshenko \& Baig(2005)]{Chernyshenko2005}
{\sc Chernyshenko, S.~I. \& Baig, M.~F.} 2005 The mechanism of streak formation
  in near-wall turbulence. {\em J. Fluid Mech.\/} {\bf 544}, 99--131.

\bibitem[Farrell \& Ioannou(1998)]{Farrell-Ioannou-1998a}
{\sc Farrell, B.~F. \& Ioannou, P.~J.} 1998 Perturbation structure and spectra
  in turbulent channel flow. {\em Theor. Comput. Fluid Dyn.\/} {\bf 11},
  215--227.

\bibitem[Farrell \& Ioannou(2012)]{Farrell2012}
{\sc Farrell, B.~F. \& Ioannou, P.~J.} 2012 Dynamics of streamwise rolls and
  streaks in turbulent wall-bounded shear flow. {\em J. Fluid Mech.\/} {\bf
  708}, 149--196.

\bibitem[Farrell \& Ioannou(2017)]{Farrell2017}
{\sc Farrell, B.~F. \& Ioannou, P.~J.} 2017 Statistical state dynamics-based
  analysis of the physical mechanisms sustaining and regulating turbulence in
  {Couette} flow. {\em Phys. Rev. Fluids\/} {\bf 2}, 084608.

\bibitem[Farrell {\em et~al.\/}(2016)Farrell, Ioannou, Jim\'enez, Constantinou,
  Lozano-Dur\'an \& Nikolaidis]{Farrell2016}
{\sc Farrell, B.~F., Ioannou, P.~J., Jim\'enez, J., Constantinou, N.~C.,
  Lozano-Dur\'an, A. \& Nikolaidis, M.-A.} 2016 A statistical state
  dynamics-based study of the structure and mechanism of large-scale motions in
  plane {Poiseuille} flow. {\em J. Fluid Mech.\/} {\bf 809}, 290--315.

\bibitem[Hack \& Moin(2018)]{Hack2018}
{\sc Hack, M. J.~P. \& Moin, P.} 2018 Coherent instability in wall-bounded
  shear. {\em J. Fluid Mech.\/} {\bf 844}, 917--955.

\bibitem[Hamilton {\em et~al.\/}(1995)Hamilton, Kim \& Waleffe]{Hamilton1995}
{\sc Hamilton, J.~M., Kim, J. \& Waleffe, F.} 1995 Regeneration mechanisms of
  near-wall turbulence structures. {\em J. Fluid Mech.\/} {\bf 287}, 317--348.

\bibitem[H{\"o}gberg {\em et~al.\/}(2003)H{\"o}gberg, Bewley \&
  Henningson]{Hogberg2003}
{\sc H{\"o}gberg, M., Bewley, T.~R. \& Henningson, D.~S.} 2003 Linear feedback
  control and estimation of transition in plane channel flow. {\em J. Fluid
  Mech.\/} {\bf 481}, 149--175.

\bibitem[Hwang \& Cossu(2010)]{Hwang2010b}
{\sc Hwang, Y. \& Cossu, C.} 2010 Linear non-normal energy amplification of
  harmonic and stochastic forcing in the turbulent channel flow. {\em J. Fluid
  Mech.\/} {\bf 664}, 51--73.

\bibitem[Jim\'enez(2012)]{Jimenez2012}
{\sc Jim{\'e}nez, J.} 2012 Cascades in wall-bounded turbulence. {\em Annu. Rev.
  Fluid Mech.\/} {\bf 44}, 27--45.

\bibitem[Jim\'enez(2013)]{Jimenez2013a}
{\sc Jim\'enez, J.} 2013 How linear is wall-bounded turbulence? {\em Phys.
  Fluids\/} {\bf 25}, 110814.

\bibitem[Jim\'enez(2018)]{Jimenez2018}
{\sc Jim\'enez, J.} 2018 Coherent structures in wall-bounded turbulence. {\em
  J. Fluid Mech.\/} {\bf 842}, P1.

\bibitem[Jim{\'e}nez \& Moin(1991)]{Jimenez1991}
{\sc Jim{\'e}nez, J. \& Moin, P.} 1991 The minimal flow unit in near-wall
  turbulence. {\em J. Fluid Mech.\/} {\bf 225}, 213--240.

\bibitem[Karp \& Cohen(2014)]{Karp2014}
{\sc Karp, M. \& Cohen, J.} 2014 Tracking stages of transition in {C}ouette
  flow analytically. {\em J. Fluid Mech.\/} {\bf 748}, 896--931.

\bibitem[Kawahara {\em et~al.\/}(2003)Kawahara, Jim\'enez, Uhlmann \&
  Pinelli]{Kawahara2003}
{\sc Kawahara, G., Jim\'enez, J., Uhlmann, M. \& Pinelli, A.} 2003 Linear
  instability of a corrugated vortex sheet -- a model for streak instability.
  {\em J. Fluid Mech.\/} {\bf 483}, 315--342.

\bibitem[Kim {\em et~al.\/}(1971)Kim, Kline \& Reynolds]{Kim1971}
{\sc Kim, H.~T., Kline, S.~J. \& Reynolds, W.~C.} 1971 The production of
  turbulence near a smooth wall in a turbulent boundary layer. {\em J. Fluid
  Mech.\/} {\bf 50}, 133--160.

\bibitem[Kim \& Lim(2000)]{Kim2000}
{\sc Kim, J. \& Lim, J.} 2000 A linear process in wall bounded turbulent shear
  flows. {\em Phys. Fluids\/} {\bf 12}, 1885--1888.

\bibitem[Kim \& Moin(1985)]{Kim1985}
{\sc Kim, J. \& Moin, P.} 1985 {Application of a fractional-step method to
  incompressible Navier-Stokes equations}. {\em J. Comp. Phys.\/} {\bf 59},
  308--323.

\bibitem[Klebanoff {\em et~al.\/}(1962)Klebanoff, Tidstrom \&
  Sargent]{Klebanoff1962}
{\sc Klebanoff, P.~S., Tidstrom, K.~D. \& Sargent, L.~M.} 1962 The
  three-dimensional nature of boundary-layer instability. {\em J. Fluid
  Mech.\/} {\bf 12}, 1--34.

\bibitem[Kline {\em et~al.\/}(1967)Kline, Reynolds, Schraub \&
  Runstadler]{Kline1967}
{\sc Kline, S.~J., Reynolds, W.~C., Schraub, F.~A. \& Runstadler, P.~W.} 1967
  The structure of turbulent boundary layers. {\em J. Fluid Mech.\/} {\bf
  30}, 741--773.

\bibitem[Landahl(1975)]{Landahl1975}
{\sc Landahl, M.~T.} 1975 Wave breakdown and turbulence. {\em SIAM J. Appl.
  Math\/} {\bf 28}, 735--756.

\bibitem[Lozano-Dur\'an \& Bae(2016)]{Lozano2016_Brief}
{\sc Lozano-Dur\'an, A. \& Bae, H.~J.} 2016 {Turbulent channel with slip
  boundaries as a benchmark for subgrid-scale models in LES}. 
  {\em Annual Research Briefs\/}, Center for Turbulence Research, pp. 97--103.

\bibitem[Lozano-Dur\'an {\em et~al.\/}(2018)Lozano-Dur\'an, Hack \&
  Moin]{Lozano2018}
{\sc Lozano-Dur\'an, A., Hack, M. J.~P. \& Moin, P.} 2018 Modeling
  boundary-layer transition in direct and large-eddy simulations using
  parabolized stability equations. {\em Phys. Rev. Fluids\/} {\bf 3}, 023901.

\bibitem[Lozano-Dur{\'a}n \& Jim{\'e}nez(2014)]{Lozano2014b}
{\sc Lozano-Dur{\'a}n, A. \& Jim{\'e}nez, J.} 2014 Time-resolved evolution of
  coherent structures in turbulent channels: characterization of eddies and
  cascades. {\em J. Fluid Mech.\/} {\bf 759}, 432--471.

\bibitem[Mansour {\em et~al.\/}(1988)Mansour, Kim \& Moin]{Mansour1988}
{\sc Mansour, N.~N., Kim, J. \& Moin, P.} 1988 Reynolds-stress and
  dissipation-rate budgets in a turbulent channel flow. {\em J. Fluid Mech.\/}
  {\bf 194}, 15--44.

\bibitem[Orlandi(2000)]{Orlandi2000}
{\sc Orlandi, P.} 2000 {\em Fluid Flow Phenomena: A Numerical Toolkit\/}.
  Springer.

\bibitem[Panton(2001)]{Panton2001}
{\sc Panton, R.~L.} 2001 Overview of the self-sustaining mechanisms of wall
  turbulence. {\em Prog. Aerosp. Sci.\/} {\bf 37}, 341--383.

\bibitem[Schmid(2007)]{Schmid2007}
{\sc Schmid, P.~J.} 2007 Nonmodal stability theory. {\em Annu. Rev. Fluid
  Mech.\/} {\bf 39}, 129--162.

\bibitem[Schoppa \& Hussain(2002)]{Schoppa2002}
{\sc Schoppa, W. \& Hussain, F.} 2002 Coherent structure generation in
  near-wall turbulence. {\em J. Fluid Mech.\/} {\bf 453}, 57--108.

\bibitem[Smits {\em et~al.\/}(2011)Smits, McKeon \& Marusic]{Smits2011}
{\sc Smits, A.~J., McKeon, B.~J. \& Marusic, I.} 2011 High-{R}eynolds number
  wall turbulence. {\em Annu. Rev. Fluid Mech.\/} {\bf 43}, 353--375.

\bibitem[Vaughan \& Zaki(2011)]{Vaughan2011}
{\sc Vaughan, N.~J. \& Zaki, T.~A.} 2011 Stability of zero-pressure-gradient
  boundary layer distorted by unsteady {Klebanoff} streaks. {\em J. Fluid
  Mech.\/} {\bf 681}, 116--153.

\bibitem[Waleffe(1997)]{Waleffe1997}
{\sc Waleffe, F.} 1997 On a self-sustaining process in shear flows. {\em Phys.
  Fluids\/} {\bf 9}, 883--900.

\bibitem[Wray(1990)]{Wray1990}
{\sc Wray, A.~A.} 1990 {Minimal-storage time advancement schemes for spectral
  methods}. {\em Tech. Rep.\/} MS 202 A-1. NASA Ames Research Center.

\end{thebibliography}

\end{document}